# Mutual Information, Relative Entropy, and Estimation in the Poisson Channel[*]

Rami Atar[1] and Tsachy Weissman[2, 1]

[1]Technion – Israel Institute of Technology, Haifa 32000, Israel
[2]Stanford University, Stanford, CA 94305

October 30, 2018

## Abstract

Let $X$ be a non-negative random variable and let the conditional distribution of a random variable $Y$, given $X$, be Poisson($\gamma \cdot X$), for a parameter $\gamma \geq 0$. We identify a natural loss function such that:

- The derivative of the mutual information between $X$ and $Y$ with respect to $\gamma$ is equal to the *minimum* mean loss in estimating $X$ based on $Y$, regardless of the distribution of $X$.

- When $X \sim P$ is estimated based on $Y$ by a mismatched estimator that would have minimized the expected loss had $X \sim Q$, the integral over all values of $\gamma$ of the excess mean loss is equal to the relative entropy between $P$ and $Q$.

For a continuous time setting where $X^T = \{X_t, 0 \leq t \leq T\}$ is a non-negative stochastic process and the conditional law of $Y^T = \{Y_t, 0 \leq t \leq T\}$, given $X^T$, is that of a non-homogeneous Poisson process with intensity function $\gamma \cdot X^T$, under the same loss function:

- The minimum mean loss in *causal* filtering when $\gamma = \gamma_0$ is equal to the expected value of the minimum mean loss in *non-causal* filtering (smoothing) achieved with a channel whose parameter $\gamma$ is uniformly distributed between 0 and $\gamma_0$. Bridging the two quantities is the mutual information between $X^T$ and $Y^T$.

- This relationship between the mean losses in causal and non-causal filtering holds also in the case where the filters employed are mismatched, i.e., optimized assuming a law on $X^T$ which is not the true one. Bridging the two quantities in this case is the sum of the mutual information and the relative entropy between the true and the mismatched distribution of $Y^T$. Thus, relative entropy quantifies the excess estimation loss due to mismatch in this setting.

These results parallel those recently found for the Gaussian channel: the I-MMSE relationship of Guo Shamai and Verdú, the relative entropy and mismatched estimation relationship of Verdú, and the relationship between causal and non-casual mismatched estimation of Weissman.

*Index Terms*- Causal estimation, Divergence, Girsanov transformation, I-MMSE, Mismatched estimation, Mutual information, Nonlinear filtering, Point process, Poisson channel, Relative entropy, Shannon theory, Statistics

# 1 Introduction

In the seminal paper [13], Guo, Shamai and Verdú discovered that the derivative of the mutual information between the input and the output in a real-valued scalar Gaussian channel, with respect to the signal-to-noise ratio (SNR), is equal to the minimum mean square error (MMSE) in estimating the input based on the output. This simple relationship holds regardless of the input distribution, and carries over essentially verbatim to vectors, as well as the continuous-time Additive White Gaussian Noise (AWGN) channel (cf. [34, 21] for even more general settings where this relationship holds). When combined with Duncan's theorem [7], it was also shown to imply a remarkable relationship between the MMSEs in causal (filtering) and non-causal (smoothing) estimation of an arbitrarily distributed continuous-time signal corrupted by Gaussian noise: the filtering MMSE at SNR level $\gamma$ is equal to the mean value of the smoothing MMSE with SNR uniformly distributed between 0 and $\gamma$. The relation of the mutual information to both types of MMSE thus served as a bridge between the two quantities.

---

[*]Research supported in part by the ISF (grants 1349/08 and 204/08), the NSF (grants CCF-1049413 and 4101-38047), the US–Israel BSF (grant 2008466), and the Technion's fund for promotion of research.

More recently, Verdú has shown in [31] that when $X \sim P$ is estimated based on $Y$ by a mismatched estimator that would have minimized the MSE had $X \sim Q$, the integral over all SNR values up to $\gamma$ of the excess MSE due to the mismatch is equal to the relative entropy between the true channel output distribution and the channel output distribution under $Q$, at SNR $= \gamma$.

This result was key in [33], where it was shown that the relationship between the causal and non-causal MMSEs continues to hold also in the mismatched case, i.e. when the filters are optimized for an underlying signal distribution that differs from the true one. The bridge between the two sides of the equality in this mismatched case was shown to be the sum of the mutual information and the relative entropy between the true and mismatched output distributions, this relative entropy thus quantifying the penalty due to mismatch.

Consider now the Poisson channel, by which we mean, for the case of scalar random variables, that $X$, the input, is a non-negative random variable while the conditional distribution of the output $Y$ given the input is given by Poisson($\gamma \cdot X$), the parameter $\gamma \geq 0$ here playing the role of SNR. In the continuous time setting, the channel input is $X^T = \{X_t, 0 \leq t \leq T\}$, a non-negative stochastic process, and conditionally on $X^T$, the output $Y^T = \{Y_t, 0 \leq t \leq T\}$ is a non-homogeneous Poisson process with intensity function $\gamma \cdot X^T$. Often referred to as the "ideal Poisson channel" [19], this model is the canonical one for describing direct detection optical communication: The channel input represents the squared magnitude of the electric field incident on the photo-detector, while its output is the counting process describing the arrival times of the photons registered by the detector. Here the energy of the channel input signal is proportional to its $l_1$ norm, rather than the $l_2$ norm as in the Gaussian channel. Thus it is the amplification factor $\gamma$ rather than $\gamma^2$ that plays the role of SNR. We refer to [32] for a review of the literature on the Poisson channel and its communication theoretic significance, and to [11] and references therein for applications of Poisson channel models in other fields.

The function $\ell_0(x) = x \log x - x + 1$, $x > 0$ (where log denotes the natural logarithm throughout), being the convex conjugate of the Poisson distribution's log moment generating function, arises naturally in analysis of Poisson and continuous time jump Markov processes in a variety of situations. These include relative entropy representation for jump Markov processes (see, e.g., equation (3.20) and Theorem 3.3 of [8]), large deviation local rate function for such processes ([8, Chapter 5 of [29]]), mutual information in the Poisson channel (Section 19.5 and equation (19.135) of [20]), and logarithmic transformations in stochastic control theory (Section 3 of [9]). It is also intimately related to change-of-measure formulae for point processes in the spirit of the Girsanov transformation (Section VI.(5.5–6) of [4], [16], [28]). It is therefore not surprising that the function $\ell_0$ appears in this paper in representations for relative entropy and related calculations. It is less obvious, however, that using it to define *estimation loss* turns out to be very useful and, in particular, gives rise to a number of results that parallel the Gaussian theory.

Enter the loss function $\ell : [0, \infty) \times [0, \infty) \to [0, \infty]$ defined by $\hat{x} \ell_0(x/\hat{x})$ or, more precisely,

$$\ell(x, \hat{x}) = x \log(x/\hat{x}) - x + \hat{x}, \tag{1}$$

where the right hand side of (1) is well-defined as an extended non-negative real number in view of our conventions $0 \log 0 = 0$, $0 \log 0/0 = 0$, $c/0 = \infty$ and $\log c/0 = \infty$ for $c > 0$. In Section 2, we exhibit properties of this loss function that show it is a natural one for measuring goodness of reconstruction of non-negative objects, and that it shares some of its key properties with the squared error loss, such as optimality of the conditional expectation under the mean loss criterion.

The goal of this paper is to show that a set of relations *identical* to those that hold for the Gaussian channel – ranging from Duncan's formula [7], to the I-MMSE of [13, 34], to Verdú's relationship between relative entropy and mismatched estimation [31], to the relationship between causal and non-causal estimation in continuous time for matched [13] and mismatched [33] filters – hold for the Poisson channel upon replacing the squared error loss by the loss function in (1).

It is instructive to note that while the relative entropy between two Gaussians of the same variance and means $m_1$ and $m_2$ is equal to $(m_1 - m_2)^2$, that between two exponentials of parameters $\lambda_1$ and $\lambda_2$ is equal to $\ell(\lambda_1, \lambda_2)$ (with additional multiplicative terms in both cases). Although this simple fact does not exclusively explain the Gaussian-Poissonian analogy, it lies at its heart, along with further properties of $\ell$ observed in Section 2.

Our emphasis is on the results for the mismatched setting, relating the cost of mismatch to relative entropy in the Poisson channel. The results for the exact (i.e., non-mismatched) setting, relating the minimum mean loss to mutual information, and causal to non-causal minimum mean estimation loss, are shown to follow as special cases. The latter results, for the exact setting, are consistent and in fact coincide with those of [14] – which considered a more general Poisson channel model that accommodates the presence of dark current – when specialized to the case of zero dark current. Our framework complements the results of [14] not only in extending the scope to the presence of mismatch, but also in highlighting that the estimation theoretic quantities obtained in [14] have minimum mean loss interpretations paralleling those from the Gaussian setting.

The remainder of the paper is organized as follows. Section 2 establishes some basic properties showing the loss function in (1) is a natural one, as discussed above. After introducing our standard notation and conventions for information measures in Section 3, we present the main results of this paper in Section 4, relating relative entropy and mismatched estimation for the Poisson channel, under the loss function in (1), for random variables and processes (causal and non-causal estimation in the latter case). In Section 5, we detail implications of our results: we show that they not only allow to recover known results from the non-mismatched Poisson channel setting such as some of those in [20] and [14], but also endow the latter with optimal estimation interpretations that allow to easily deduce results paralleling those that have been established in the Gaussian setting. We present additional consequences, such as an estimation theoretic representation of entropy, the relationship between causal and non-causal (matched and mismatched) estimation in continuous time, and some of its implications. Section 6 illustrates some of our key findings via a couple of simplistic examples where the underlying noise-free process is a DC signal. Section 7 is dedicated to proving our results. We end that section by indicating how some of the results carry over to accommodate the presence of feedback. An alternative route to proving the main results via a more elementary analysis is described in Section 8. We conclude in Section 9 with a summary of our findings and some related future directions.

## 2 A Natural Loss Function

Throughout the paper we use $\ell$ to denote the loss function specified in (1). In this section, we collect a couple of lemmas suggesting that $\ell$ is a natural function for measuring goodness of estimation of a non-negative random variable, and that it possesses some of the celebrated properties of squared error loss that make the latter so popular.

**Lemma 2.1** *The function $\ell$ defined in (1) has the following properties:*

1. *Non-negativity: $\ell(x, \hat{x}) \geq 0$ with equality if and only if $x = \hat{x}$.*

2. *Convexity: $\ell(x, \hat{x})$ is convex in each of its arguments.*

3. *Scaling: $\ell(\alpha x, \alpha \hat{x}) = \alpha \cdot \ell(x, \hat{x})$, $\alpha \geq 0$.*

4. *Unboundedness of loss for underestimation: For any $x > 0$, $\lim_{\hat{x} \to 0^+} \ell(x, \hat{x}) = \ell(x, 0) = \infty$.*

One way of seeing why the first (non-negativity) property stated in the lemma holds is to recall that

$$\ell(x, \hat{x}) = \hat{x} \cdot \ell_0(x/\hat{x}) \tag{2}$$

and note that the function

$$\ell_0(x) = x \log x - x + 1 \tag{3}$$

is a (convex) non-negative function assuming its unique minimum value of 0 at $x = 1$, cf. Figure 1-(b). We leave the elementary verification of the remaining properties collected in the lemma to the reader. Note that the first two properties in the lemma are enough to imply unboundedness of the loss for overestimation, i.e., that for any $x \geq 0$, $\lim_{\hat{x} \to \infty} \ell(x, \hat{x}) = \infty$. The fourth property shows that the same holds true for underestimation, a pleasing property when measuring goodness of reconstruction of non-negative quantities, not possessed by the more common loss functions such as absolute or squared error.

To note another key property, recall the squared error loss function

$$\ell_{SE}(x, \hat{x}) = (x - \hat{x})^2, \tag{4}$$

which satisfies, for any random variable $X$ of finite variance,

$$E\left[\ell_{SE}(X,\hat{x})\right] = E\left[\ell_{SE}(X,EX)\right] + \ell_{SE}(EX,\hat{x}). \tag{5}$$

The same relationship holds under our present loss function as well:

**Lemma 2.2** *For any non-negative random variable $X$ with $E[X\log^+ X] < \infty$, and any $\hat{x} \in [0,\infty)$,*

$$E\left[\ell(X,\hat{x})\right] = E\left[\ell(X,EX)\right] + \ell(EX,\hat{x}). \tag{6}$$

*Proof:* When $X = 0$ a.s., $EX = 0$, and this identity follows from our conventions. Otherwise $EX > 0$ and

$$E\left[\ell(X,\hat{x})\right] = E\left[X\log\frac{X}{\hat{x}} - X + \hat{x}\right] \tag{7}$$

$$= E\left[X\log\left(\frac{X}{\hat{x}}\frac{EX}{EX}\right) - X + \hat{x}\right] \tag{8}$$

$$= E\left[X\log\frac{X}{EX} - X + EX\right] + \left[EX\log\frac{EX}{\hat{x}} - EX + \hat{x}\right] \tag{9}$$

$$= E\left[\ell(X,EX)\right] + \ell(EX,\hat{x}). \quad \Box \tag{10}$$

One immediate consequence of (6), when put together with the first (non-negativity) property in Lemma 2.1, is the fact that $EX$ uniquely minimizes $E\left[\ell(X,\hat{x})\right]$ over all $\hat{x}$:

$$\min_{\hat{x}} E\left[\ell(X,\hat{x})\right] = E\left[\ell(X,EX)\right] = E[X\log X] - E[X]\log E[X], \tag{11}$$

and thus $E[X\log X] - E[X]\log E[X]$ plays a role analogous to that played by variance under squared error loss. An immediate consequence of (11) is that conditional expectation $E[X|Y]$ is the unique estimator of $X$ based on $Y$ minimizing the mean loss not only under $\ell_{SE}$, but also under $\ell$. This property is in fact common to (and characterizes) the family of Bregman loss functions [1].

Another key property shared by the loss function $\ell$ and squared error loss, exhibited respectively in (6) and (5), is that, beyond quantifying the loss, it also quantifies the price of mismatch, i.e., the excess loss due to using $\hat{x}$ in lieu of $EX$ (cf., respectively, the second terms on the right hand sides of (6) and (5)). This property, which in the case of squared error loss is due to orthogonality, is as key in the derivation of our results as the orthogonality principle is for deriving the results in [7, 13, 31, 33].

## 3  Notation and Conventions

Our conventions and notation for information measures, such as mutual information and relative entropy, are standard. The initiated reader is advised to skip this section. If $U, V, W$ are three random variables taking values in Polish spaces $\mathcal{U}, \mathcal{V}, \mathcal{W}$, respectively, and defined on a common probability space with a probability measure $P$, we let $P_U, P_{U,V}$ etc. denote the probability measures induced on $\mathcal{U}$, the pair $(\mathcal{U}, \mathcal{V})$ etc. while e.g., $P_{U|V}$ denotes a regular version of the conditional distribution of $U$ given $V$. $P_{U|v}$ is the distribution on $\mathcal{U}$ obtained by evaluating that regular version at $v$. If $Q$ is another probability measure on the same measurable space we similarly denote $Q_U, Q_{U|V}$, etc. As usual, given two measures on the same measurable space, e.g., $P$ and $Q$, define their relative entropy (divergence) by

$$D(P\|Q) = \int\left[\log\frac{dP}{dQ}\right]dP \tag{12}$$

when $P$ is absolutely continuous w.r.t. $Q$, defining $D(P\|Q) = \infty$ otherwise. An immediate consequence of the definitions of relative entropy and of the Radon-Nikodym derivative is that if $f : \mathcal{U} \to \mathcal{V}$ is measurable and one-to-one, and $V = f(U)$, then

$$D(P_U\|Q_U) = D(P_V\|Q_V). \tag{13}$$

Following [5], we further use the notation

$$D(P_{U|V}\|Q_{U|V}|P_V) = \int D(P_{U|v}\|Q_{U|v})dP_V(v), \tag{14}$$

where on the right side $D(P_{U|v}\|Q_{U|v})$ is a divergence in the sense of (12) between the measures $P_{U|v}$ and $Q_{U|v}$. It will be convenient to write

$$D(P_{U|V}\|Q_{U|V}) \qquad (15)$$

to denote $f(V)$ when $f(v) = D(P_{U|v}\|Q_{U|v})$. Thus $D(P_{U|V}\|Q_{U|V})$ is a random variable while $D(P_{U|V}\|Q_{U|V}|P_V)$ is its expectation under $P$. With this notation, the chain rule for relative entropy (cf., e.g., [6, Subsection D.3]) is

$$D(P_{U,V}\|Q_{U,V}) = D(P_U\|Q_U) + D(P_{V|U}\|Q_{V|U}|P_U) \qquad (16)$$

and is valid regardless of the finiteness of both sides of the equation.

The mutual information between $U$ and $V$ is defined as

$$I(U;V) = D(P_{U,V}\|P_U \times P_V), \qquad (17)$$

where $P_U \times P_V$ denotes the product measure induced by $P_U$ and $P_V$. We note in passing, in line with the comment on relative entropy and one-to-one transformations leading to (13), that if $f$ and $g$ are two measurable one-to-one transformations and $A = f(U)$ while $B = g(V)$, then

$$I(U;V) = I(A;B). \qquad (18)$$

Finally, the conditional mutual information between $U$ and $V$, given $W$, is defined as

$$I(U;V|W) = D(P_{U,V|W}\|P_{U|W} \times P_{V|W}|P_W). \qquad (19)$$

The roles of $U, V, W$ will be played in what follows by scalar random variables, vectors, or processes.

## 4 Relative Entropy and Mismatched Estimation

### 4.1 Random Variables

Suppose that $X$ is a non-negative random variable and the conditional law of a r.v. $Y_\gamma$, given $X$, is Poisson($\gamma X$). If $X \sim P$, denote expectation w.r.t. the corresponding joint law of $X$ and $Y_\gamma$ by $E_P$, the distribution of $Y_\gamma$ by $P_{Y_\gamma}$, the conditional expectation by $E_P[X|Y_\gamma]$, etc. We denote the mutual information by $I_P(X;Y_\gamma)$ or simply $I(X;Y_\gamma)$ when there is no ambiguity. Let further $\mathsf{mle}_{P,Q}(\gamma)$ denote the mean loss under $\ell$ in estimating $X$ based on $Y_\gamma$ using the estimator that would have been optimal had $X \sim Q$ when in fact $X \sim P$, i.e.,

$$\mathsf{mle}_{P,Q}(\gamma) \triangleq E_P\left[\ell\big(X, E_Q[X|Y_\gamma]\big)\right]. \qquad (20)$$

The following is a new representation of relative entropy, paralleling the Gaussian channel result of [31]:

**Theorem 4.1** *For any pair $P, Q$ of probability measures over $[a,b]$, where $0 < a < b < \infty$,*

$$D(P\|Q) = \int_0^\infty [\mathsf{mle}_{P,Q}(\gamma) - \mathsf{mle}_{P,P}(\gamma)]\, d\gamma \qquad (21)$$

Theorem 4.1 is a direct consequence of the fact (proved in Section 7) that

$$\lim_{\gamma \to \infty} D(P_{Y_\gamma}\|Q_{Y_\gamma}) = D(P\|Q), \qquad (22)$$

combined with the following result, which is the Poisson parallel of [31, Equation (24)]:

**Theorem 4.2** *For any $P, Q$ as in Theorem 4.1, and for any $\gamma \geq 0$,*

$$D(P_{Y_\gamma}\|Q_{Y_\gamma}) = \int_0^\gamma [\mathsf{mle}_{P,Q}(\alpha) - \mathsf{mle}_{P,P}(\alpha)]\, d\alpha. \qquad (23)$$

To note one immediate implication of Theorem 4.2, the non-negativity of the integrand on the right hand side of (23), as follows from (6), implies that $D(P_{Y_\gamma}\|Q_{Y_\gamma})$ increases with $\gamma$. Additional implications are pointed out in Section 5.

## 4.2 Continuous-Time Stochastic Processes

Fix $T > 0$. Denote by $\mathbb{D}$ the space of right-continuous paths with left limits from $[0, T]$ to $\mathbb{R}$. Endow $\mathbb{D}$ with the usual Skorohod topology [3] and denote by $\mathcal{D}$ the Borel $\sigma$-algebra of $\mathbb{D}$. Denote by $\mathcal{P}$ the collection of probability measures $P$ on $(\mathbb{D}, \mathcal{D})$ under which for $P$-a.e. $\xi \in \mathbb{D}$, $\xi$ is bounded between two positive constants.

A measurable space $(\Omega, \mathcal{F})$ is given, on which a stochastic process $X^T = \{X_t, 0 \leq t \leq T\}$ and, for each $\gamma > 0$, a stochastic process $Y_\gamma^T = \{Y_{\gamma,t}, 0 \leq t \leq T\}$, are given. These processes represent the signal and observation, respectively. The sample paths of each of them are in $\mathbb{D}$, and each is assumed to be measurable as a map from $(\Omega \times [0, T], \mathcal{F} \otimes \mathcal{B}([0, T]))$ to $(\mathbb{R}_+, \mathcal{B}(\mathbb{R}_+))$ (where throughout, $\mathcal{B}(U)$ denotes the Borel $\sigma$-algebra of a metric space $U$). Given $\gamma \geq 0$, we are interested in probability measures $\hat{P}$ on $(\Omega, \mathcal{F})$ under which the measure induced by $X$ on $(\mathbb{D}, \mathcal{D})$ is in $\mathcal{P}$, and $Y_\gamma^T$ is jointly distributed with $X^T$ in such a way that, given $X^T$, $Y_\gamma^T$ is a non-homogenous Poisson process with intensity function $\gamma \cdot X^T$. We denote by $\hat{\mathcal{P}}_\gamma = \hat{\mathcal{P}}_\gamma(\Omega, \mathcal{F})$ the collection of all such measures.

Given $\gamma \geq 0$ and a probability measure $P$ on $(\mathbb{D}, \mathcal{D})$ pick $\hat{P} \in \hat{\mathcal{P}}_\gamma$ for which the law of $X^T$ on $\mathbb{D}$ is $P$, and let $E_P$ denote expectation w.r.t. the joint law of $X^T$ and $Y_\gamma^T$ on $\mathbb{D}^2$ under $\hat{P}$. Although this law depends on $\gamma$, there is no need to indicate $\gamma$ in the notation $E_P$ because $Y_\gamma^T$ itself depends on $\gamma$. Let $\mathsf{mle}_{P,Q}(\gamma)$ denote the expected cumulative loss in *non-causal* filtering of $X^T$ based on $Y_\gamma^T$ when $X^T \sim P$ but the filter used is optimized for $X^T \sim Q$, i.e.,

$$\mathsf{mle}_{P,Q}(\gamma) \triangleq E_P \left[ \int_0^T \ell \left( X_t, E_Q[X_t | Y_\gamma^T] \right) dt \right]. \tag{24}$$

Note that the definitions in (20) and (24) are consistent with each other, and which of them applies is dictated unambiguously according to whether $P$ and $Q$ govern random variables or processes. Theorem 4.2 carries over to stochastic processes as follows:

**Theorem 4.3** *Let $P$ and $Q$ be two probability measures that are members of $\mathcal{P}$. For $\gamma \geq 0$,*

$$D(P_{Y_\gamma^T} \| Q_{Y_\gamma^T}) = \int_0^\gamma [\mathsf{mle}_{P,Q}(\alpha) - \mathsf{mle}_{P,P}(\alpha)] \, d\alpha. \tag{25}$$

Let now $\mathsf{cmle}_{P,Q}(\gamma)$ denote the expected cumulative loss in *causal* filtering of $X^T$ based on $Y_\gamma^T$ when $X^T \sim P$ but the filter used is optimized for $X^T \sim Q$, i.e.,

$$\mathsf{cmle}_{P,Q}(\gamma) \triangleq E_P \left[ \int_0^T \ell \left( X_t, E_Q[X_t | Y_\gamma^t] \right) dt \right]. \tag{26}$$

The right hand side of (24) and that of (26) differ only in that the conditional expectation appearing in the former has the entire process $Y_\gamma^T$ in the conditioning, while in the latter only $Y_\gamma^t$, the process up to time $t$. Our main result regarding $\mathsf{cmle}_{P,Q}(\gamma)$ is that its excess value above the mean filtering loss of the optimal causal filter is proportional to the relative entropy between $P_{Y_\gamma^T}$ and $Q_{Y_\gamma^T}$:

**Theorem 4.4** *Let $P$ and $Q$ be two probability measures that are members of $\mathcal{P}$. For $\gamma \geq 0$,*

$$D(P_{Y_\gamma^T} \| Q_{Y_\gamma^T}) = \gamma \cdot [\mathsf{cmle}_{P,Q}(\gamma) - \mathsf{cmle}_{P,P}(\gamma)]. \tag{27}$$

Put together, Theorem 4.3 and Theorem 4.4 yield, for $\gamma > 0$,

$$\mathsf{cmle}_{P,Q}(\gamma) - \mathsf{cmle}_{P,P}(\gamma) = \frac{1}{\gamma} \int_0^\gamma [\mathsf{mle}_{P,Q}(\alpha) - \mathsf{mle}_{P,P}(\alpha)] \, d\alpha = \frac{1}{\gamma} D(P_{Y_\gamma^T} \| Q_{Y_\gamma^T}), \tag{28}$$

which is the Poissonian analogue of [33, Theorem 2]. On a technical note, the r.h.s. of (24), (25) and (26) are well-defined as integrals of non-negative Borel measurable functions, as will follow from our treatment in Section 7.

## 5 Implications

### 5.1 Mutual Information and Minimum Mean Estimation Loss

Let $X$ be a non-negative random variable and, for $\gamma > 0$, let $Y_\gamma$ be a non-negative integer-valued random variable, jointly distributed with $X$ such that the conditional law of $Y_\gamma$ given $X$ is $\text{Poisson}(\gamma X)$. When specialized to this setting, Theorem 2 of [14] gives

$$\frac{d}{d\gamma}I(X;Y_\gamma) = E\left[X \log X - E[X|Y_\gamma] \log E[X|Y_\gamma]\right]. \tag{29}$$

It is instructive to observe that the right hand side of (29) is nothing but the minimum mean loss in estimating $X$ based on $Y_\gamma$ under the loss function $\ell$. Indeed, denoting this minimum mean loss by $\mathsf{mmle}(\gamma)$, i.e.,

$$\mathsf{mmle}(\gamma) \triangleq E\left[\ell\left(X, E[X|Y_\gamma]\right)\right], \tag{30}$$

we have

$$E\left[\ell\left(X, E[X|Y_\gamma]\right)\right] = E\left[X \log \frac{X}{E[X|Y_\gamma]} - X + E[X|Y_\gamma]\right] \tag{31}$$

$$= E\left[X \log X - X \log E[X|Y_\gamma]\right] \tag{32}$$

$$= E\left[X \log X - E[X|Y_\gamma] \log E[X|Y_\gamma]\right]. \tag{33}$$

Thus, (29) can be stated as the "I-MMLE" relationship

$$\frac{d}{d\gamma}I(X;Y_\gamma) = \mathsf{mmle}(\gamma), \tag{34}$$

in complete analogy with the I-MMSE relationship of [13]. To see one immediate benefit of this realization that the right hand side of (29) coincides with the minimum mean loss in the right hand side of (34), we first go through the following data processing argument: Fix $\gamma' < \gamma$, let $\{B_i\}_{i\geq 1}$ be i.i.d. Bernoulli($\gamma'/\gamma$) independent of $(X, Y_\gamma)$, and note that $\left(X, \sum_{i=1}^{Y_\gamma} B_i\right)$ is equal in distribution to $(X, Y_{\gamma'})$. Since estimating $X$ based on $\sum_{i=1}^{Y_\gamma} B_i$, which is a function of $Y_\gamma$ and the randomization sequence $\{B_i\}$, cannot be better (in the sense of minimizing the expected loss under $\ell$) than estimating $X$ based on $Y_\gamma$, we have $\mathsf{mmle}(\gamma') \geq \mathsf{mmle}(\gamma)$. Thus, $\mathsf{mmle}(\gamma)$ is non-increasing with $\gamma$ which, when combined with (34), yields the following analogue of [13, Corollary 1]:

**Corollary 5.1** $I(X;Y_\gamma)$ is concave in $\gamma$.

It is also worth pointing out that the I-MMLE relationship can be viewed as a direct consequence of Theorem 4.2. Indeed, in the notation of Section 4.1, (34) is expressed as

$$\frac{d}{d\gamma}I_P(X;Y_\gamma) = \mathsf{mle}_{P,P}(\gamma), \tag{35}$$

which can be inferred from:

$$I_P(X;Y_\gamma) = \int D(P_{Y_\gamma|X=x} \| P_{Y_\gamma}) dP_X(x) \tag{36}$$

$$\stackrel{(a)}{=} \int \left[\int_0^\gamma [\mathsf{mle}_{\delta_x, P}(\alpha) - \mathsf{mle}_{\delta_x, \delta_x}(\alpha)] d\alpha\right] dP_X(x) \tag{37}$$

$$= \int \left[\int_0^\gamma E_{\delta_x}\left[\ell(x, E_P[X|Y_\alpha])\right] d\alpha\right] dP_X(x) \tag{38}$$

$$= \int_0^\gamma E_P\left[\ell(X, E_P[X|Y_\alpha])\right] d\alpha \tag{39}$$

$$= \int_0^\gamma \mathsf{mle}_{P,P}(\alpha) d\alpha, \tag{40}$$

where we use $\delta_x$ to denote the degenerate distribution on $x$ and $(a)$ follows by applying Theorem 4.2 on the integrand in (36). Note that (40) is nothing but the integral version of (35).

A similar exercise – of expressing the mutual information between the channel input and output as a relative entropy between the distribution of the output conditioned on a particular channel input and the unconditioned channel output distribution, integrated over the channel input distribution, and using the relevant relationship from Section 4 to express the integrand – can be performed in the continuous-time setting of Section 4.2. Indeed, application of Theorem 4.3 on the said integrand gives

$$I_P\left(X^T; Y_\gamma^T\right) = \int_0^\gamma \mathsf{mle}_{P,P}(\alpha) d\alpha. \tag{41}$$

The relationship (41) is consistent with [14, Theorem 4], the two in fact coinciding when specializing the latter to zero dark current. In a similar way, application of Theorem 4.4 gives

$$I_P\left(X^T; Y_\gamma^T\right) = \gamma \cdot \mathsf{cmle}_{P,P}(\gamma), \tag{42}$$

which, as in (33), is seen to be equivalent to the known relationship

$$I_P(X^T; Y_\gamma^T) = \gamma \cdot E_P\left[\int_0^T \left(X_t \log X_t - E_P[X_t|Y_\gamma^t] \log E_P[X_t|Y_\gamma^t]\right) dt\right], \tag{43}$$

cf., e.g., [20, Subection 19.5]. The representation (42) highlights the optimal estimation interpretation of this relationship and, through that, the close analogy with Duncan's theorem [7, Theorem 3].

Finally, going back to the setting of the "I-MMLE" relationship for scalar random variables, we note that the conditional entropy of the output given the input is given by

$$H(Y_\gamma|X) = E\left[\gamma X \cdot (1 - \log(\gamma X)) + e^{-\gamma X} \sum_{k=0}^\infty \frac{(\gamma X)^k \log(k!)}{k!}\right], \tag{44}$$

so, in particular, is dependent on both $\gamma$ and the distribution of $X$, in contrast to the Gaussian channel setting of [13] where the (differential) entropy of the output given the input depends on neither. Thus, while the mutual information can be replaced by the (differential) entropy of the channel output in the I-MMSE relationship of [13], this is not the case for the I-MMLE relationship of our present setting, which further consolidates the role of mutual information rather than channel output entropy as key in the relations between information and estimation.

## 5.2 Estimation Theoretic Representation of Entropy

For a random variable $X$ taking values in a countable set $\mathcal{A}$, and a mapping $g : \mathcal{A} \to [0, \infty)$, let $Z_\gamma$ be jointly distributed with $X$ such that the conditional law of $Z_\gamma$ given $X$ is Poisson$(\gamma \cdot g(X))$. With benign abuse of notation, we write Poisson$(\gamma \cdot g(X))$ in lieu of $Z_\gamma$ when convenient. The following is a direct consequence of Theorem 4.1.

**Corollary 5.2** *For any discrete random variable $X \sim P$ supported on the alphabet $\mathcal{A}$, for any one-to-one mapping $g : \mathcal{A} \to [0, \infty)$, and for any $x \in \mathcal{A}$ with $P(X = x) > 0$, denoting expectation w.r.t. $P$ by $E$,*

$$\log \frac{1}{P(X=x)} = \int_0^\infty E\left[\ell\big(g(X), E\left[g(X)|\text{Poisson}(\gamma \cdot g(X))\right]\big)|X = x\right] d\gamma. \tag{45}$$

*Proof:* Apply Theorem 4.1 with $P = \delta_{g(x)}$ and an arbitrary $Q$ supported on $g(\mathcal{A})$ to obtain

$$\log \frac{1}{Q(X=x)} = \log \frac{1}{Q(g(X)=g(x))} = \int_0^\infty E\left[\ell\big(g(x), E_Q\left[g(X)|Poisson(\gamma \cdot g(X))\right]\big)|X = x\right] d\gamma, \tag{46}$$

and note that (45) is nothing but (46) specialized to $Q = P$. □

Averaging over (45) with respect of $P$ gives the following analogue to Theorem 13 of [13]:

**Corollary 5.3** *For any discrete random variable $X$ taking values in $\mathcal{A}$, and for any one-to-one mapping $g : \mathcal{A} \to [0, \infty)$,*

$$H(X) = \int_0^\infty E\left[\ell\big(g(X), E\left[g(X)|\text{Poisson}(\gamma \cdot g(X))\right]\big)\right] d\gamma. \tag{47}$$

Corollary 5.3 can also be deduced directly from (40) by noting that, for discrete $X$, $I(X; Y_\gamma) \to H(X)$ as $\gamma \to \infty$, and that if the discrete $X$ takes values in the alphabet $\mathcal{A}$ then $H(X) = H(g(X))$ for any one-to-one mapping $g : \mathcal{A} \to \mathbb{R}$. It is interesting that the integrals on the right-hand sides of (45) and (47) do not depend on $g$, a fact that seems hard to deduce directly from estimation theoretic considerations.

## 5.3 Relationship Between Causal and Non-Causal Estimation and the Price of Mismatch in Continuous Time

Our main findings for the continuous-time setting are summarized in the theorem that follows, relating the causal estimation error, the non-causal estimation error, the mutual information, and the relative entropy. It follows directly from combining (28), (41) and (42).

**Theorem 5.5** *Let $P$ and $Q$ be two probability laws on the non-negative process $X^T$ that are members of $\mathcal{P}$. For any $\gamma > 0$,*

$$\mathsf{cmle}_{P,Q}(\gamma) = \frac{1}{\gamma} \int_0^\gamma \mathsf{mle}_{P,Q}(\alpha) d\alpha = \frac{1}{\gamma} \left[ I_P(X^T; Y_\gamma^T) + D(P_{Y_\gamma^T} \| Q_{Y_\gamma^T}) \right]. \tag{48}$$

We list a few observations that are implied for the continuous-time setting by Theorem 5.5 in a manner similar to that in which the observations in Section 2 of [33] are implied by the main result therein. We refer to that section for the details.

- $\mathsf{cmle}_{P,Q}(\gamma)$ *and* $\mathsf{mle}_{P,Q}(\gamma)$ *do not necessarily decrease with* $\gamma$: In the mismatched case, when $P \neq Q$, $\mathsf{cmle}_{P,Q}(\gamma)$ and $\mathsf{mle}_{P,Q}(\gamma)$ need not decrease with $\gamma$, nor need the relationship $\mathsf{cmle}_{P,Q}(\gamma) \geq \mathsf{mle}_{P,Q}(\gamma)$ hold. The two properties in fact determine each other:

$$\mathsf{mle}_{P,Q}(\gamma) > \mathsf{cmle}_{P,Q}(\gamma) \quad \text{if and only if} \quad \frac{d}{d\gamma} \mathsf{cmle}_{P,Q}(\gamma) > 0. \tag{49}$$

In words: an increase in SNR deteriorates the mismatched causal estimation performance if and only if the latter is better than the non-causal estimation performance. We give an extreme example of such a scenario in Section 6.2.

- *Invariance of the Mismatched Filtering Performance to the Direction of Time:* Let $\mathsf{acmle}_{P,Q}(\gamma)$ denote the mean estimation loss achieved by the anti-causal mismatched filter, i.e.,

$$\mathsf{acmle}_{P,Q}(\gamma) \triangleq E_P \left[ \int_0^T \ell\big(X_t, E_Q[X_t | \{Y_{\gamma,s} - Y_{\gamma,t}\}_{t \leq s \leq T}]\big) dt \right]. \tag{50}$$

By the invariance of the mutual information and of the relative entropy in (48) to the direction of the flow of time (apply, respectively, (18) and (13) with the role of the transformations $f$ and $g$ played by time reversal), we obtain

$$\mathsf{acmle}_{P,Q}(\gamma) = \mathsf{cmle}_{P,Q}(\gamma). \tag{51}$$

It is remarkable that (51) holds under essentially no restrictions on $P$, on $Q$, or on the relationship between them.

- *Factor of 2 Relationship at Low SNR:* Assuming that at $\gamma = 0$ $\mathsf{cmle}_{P,Q}(\gamma)$ is continuously differentiable and non-zero, we have

$$\lim_{\gamma \to 0} \frac{\int_0^T E_P[\ell(X_t, E_Q X_t)] dt - \mathsf{mle}_{P,Q}(\gamma)}{\int_0^T E_P[\ell(X_t, E_Q X_t)] dt - \mathsf{cmle}_{P,Q}(\gamma)} = 2, \tag{52}$$

i.e., the non-causal error approaches its low SNR limit twice as rapidly as the causal error.

- *High SNR Behavior of $D(P_{Y_\gamma^T} \| Q_{Y_\gamma^T})$:* Assume the relationship between $P$ and $Q$ is sufficiently regular to imply

$$\lim_{\gamma \to \infty} \mathsf{cmle}_{P,Q}(\gamma) = \lim_{\gamma \to \infty} \mathsf{mle}_{P,Q}(\gamma) = 0. \tag{53}$$

$D(P_{Y_\gamma^T} \| Q_{Y_\gamma^T})$ exhibits one of the following possible behaviors in the high SNR regime:

1. $D(P_{Y_\gamma^T} \| Q_{Y_\gamma^T}) \equiv 0$ for all $\gamma > 0$. This can happen if and only if $D(P\|Q) = 0$, i.e., the non-mismatched setting.
2. $D(P_{Y_\gamma^T} \| Q_{Y_\gamma^T}) = \Theta(1)$, which can happen if and only if $0 < D(P\|Q) < \infty$.

3. $\lim_{\gamma \to \infty} D(P_{Y^T_\gamma} \| Q_{Y^T_\gamma}) = \infty$ but $D(P_{Y^T_\gamma} \| Q_{Y^T_\gamma}) = o(\gamma)$, which can happen if and only if $D(P\|Q) = \infty$. I.e., when $D(P\|Q) = \infty$, $D(P_{Y^T_\gamma} \| Q_{Y^T_\gamma})$ increases without bound with increasing SNR, but sub-linearly.

- *"Semi-Stochastic" Setting:* Suppose that $X_t = \lambda_t$, $0 \le t \le T$, $\lambda_t \ge 0$ being a deterministic signal. Applying Theorem 5.5 with $P$ degenerate on $\lambda^T$ gives

$$E\left[\int_0^T \ell\left(\lambda_t, E_Q[X_t | Y_\gamma^t]\right) dt\right] = \frac{1}{\gamma} \int_0^\gamma E\left[\int_0^T \ell\left(\lambda_t, E_Q[X_t | Y_\alpha^T]\right) dt\right] d\alpha = \frac{1}{\gamma} D(P_{Y^T_\gamma} \| Q_{Y^T_\gamma}), \tag{54}$$

where here $P_{Y^T_\gamma}$ is the law of a non-homogeneous Poisson process with intensity function $\gamma \cdot \lambda^T$, and the expectations are with respect to this measure. The relationship (54) can be thought of as the non-Bayesian version of (48).

We refer to Section 2 of [33] for the details leading to the above observations, as well as additional observations and results that, equipped with Theorem 5.5, carry over verbatim from the Gaussian to the Poisson channel, such as the structure and performance of minimax causal estimators and their direct relation to minimax source coding via redundancy-capacity theory [10, 27, 25].

## 6 Example: A DC Signal

We now work out two examples. In both, the underlying noise-free process is a DC signal known to be such by the mismatched filter, the mismatch being only in the prior distribution on its amplitude. As simplistic as this scenario may be, it illustrates how some of the key observations made above play themselves out in concrete cases.

### 6.1 Binary DC Signal

Consider the case where $X^T$ is a binary DC process, i.e., $X_t \equiv X$, where $X$ takes the values 1 and 0 with probabilities $p$ and $1-p$, and without loss of generality take $T = 1$. Suppose that the mismatched filter is designed knowing that $X^T$ is a binary DC process, but under the assumption that $X$ takes the values 1 and 0 with probabilities $q$ and $1-q$. Subscripting with $p$ and $q$ to denote the respective measures, we have

$$E_p[X_t | Y_\gamma^t = y] = \begin{cases} 1 & \text{if } y \ge 1 \\ \frac{pe^{-\gamma t}}{1-p+pe^{-\gamma t}} & \text{if } y = 0 \end{cases} \tag{55}$$

and

$$E_p\left[\ell\left(X_t, E_q[X_t | Y_\gamma^t]\right)\right] = g(p, q, \gamma t), \tag{56}$$

where

$$g(p, q, \gamma) = \frac{qe^{-\gamma}(1-p)}{1-q+qe^{-\gamma}} + \left[\log \frac{1-q+qe^{-\gamma}}{qe^{-\gamma}} - \frac{1-q}{1-q+qe^{-\gamma}}\right] pe^{-\gamma}. \tag{57}$$

Thus

$$\mathsf{cmle}_{P,Q}(\gamma) = \int_0^1 g(p,q,\gamma t) dt = \frac{1}{\gamma} e^{-\gamma} \left\{-p \log\left[1 + e^\gamma \left(\frac{1}{q}-1\right)\right] + e^\gamma \left(p \log \frac{1}{q} + (p-1) \log\left[1 + \left(e^{-\gamma}-1\right)q\right]\right)\right\} \tag{58}$$

while

$$\mathsf{mle}_{P,Q}(\gamma) = g(p, q, \gamma). \tag{59}$$

The curves $\mathsf{cmle}_{P,Q}(\gamma)$ and $\mathsf{mle}_{P,Q}(\gamma)$ are plotted in Figure 2, along with those for the exact setting $\mathsf{cmle}_{P,P}(\gamma)$ and $\mathsf{mle}_{P,P}(\gamma)$. Theorem 5.5 implies that the area of the dark rectangle is equal to the area under the curve of $\mathsf{mle}_{P,P}(\gamma)$, which are both equal to the mutual information between the clean and noisy signal. It further implies that the sum of the areas of the dark and the light colored rectangles is equal to the area under the curve $\mathsf{mle}_{P,Q}(\gamma)$, and both are equal to the mutual information plus the relative entropy between the true and the mismatched channel output distribution. Thus, the area of the light colored rectangle is equal to the relative entropy. Also evident in the figure is the 'factor of 2' relationship of Equation (52), which holds for the two pairs of curves (mismatched and exact).

## 6.2 Support of $P$ Not in Support of $Q$

In the previous example, $P$ and $Q$ had the same support, which guarantees $\lim_{\gamma \to \infty} \mathsf{mle}_{P,Q}(\gamma) = \lim_{\gamma \to \infty} \mathsf{cmle}_{P,Q}(\gamma) = 0$ despite the mismatch. Also, in that example $\mathsf{mle}_{P,Q}(\gamma)$ and $\mathsf{cmle}_{P,Q}(\gamma)$ were monotonically decreasing with $\gamma$. In general, neither of these properties need to hold, as was mentioned in Section 5.3 and is illustrated in the following extreme example.

Suppose that, under $P$, the signal is deterministic and constant at $1/2$, i.e., $X_t \equiv 1/2$ for all $0 \leq t \leq T = 1$. Under $Q$, $X^T$ is a binary DC process as in the previous example with $q = 1/2$. Using (55), an elementary calculation yields

$$E_P\left[\ell\left(X_t, E_Q[X_t|Y_\gamma^t]\right)\right] = f(\gamma t), \tag{60}$$

where

$$f(\gamma) = \ell(1/2, 1) \cdot (1 - e^{-\gamma/2}) + \ell\left(1/2, \frac{e^{-\gamma}}{1 + e^{-\gamma}}\right) \cdot e^{-\gamma/2}. \tag{61}$$

Thus

$$\mathsf{mle}_{P,Q}(\gamma) = f(\gamma) \tag{62}$$

while elementary manipulations give

$$\mathsf{cmle}_{P,Q}(\gamma) = \int_0^1 f(\gamma t) dt \tag{63}$$

$$= -\frac{1}{24\gamma}\left[-24 + 24e^{-\gamma/2} + 3\gamma^2 + \pi^2 - 24 \cdot \text{Gudermannian}\left(-\frac{\gamma}{2}\right) + \gamma 12 \log 2 - 24 \log 2 \right. \tag{64}$$

$$\left. + e^{-\gamma/2} 24 \log 2 + 12\gamma \log[1 + e^\gamma] - 12\gamma \log\left[e^\gamma \text{Cosh}\left(\frac{\gamma}{2}\right)\right] + 12 \cdot \text{Li}_2\left(-e^\gamma\right)\right], \tag{65}$$

where $\text{Gudermannian}(x) = 2\arctan(e^x) - \pi/2$ and $\text{Li}_2(x) = \sum_{k=1}^\infty x^k/k^2$. Figure 3 displays the curves of $\mathsf{mle}_{P,Q}(\gamma)$ and $\mathsf{cmle}_{P,Q}(\gamma)$ in (62) and (65). In accord with Theorem 5.5, the area of the rectangle is equal to the area under the curve of $\mathsf{mle}_{P,Q}(\gamma)$, which are both equal to the relative entropy between the true and the mismatched channel output distribution, noting that the mutual information in this case is zero since the clean signal is deterministic.

This example shows that $\mathsf{mle}_{P,Q}(\gamma)$ and $\mathsf{cmle}_{P,Q}(\gamma)$ need not vanish with increasing $\gamma$, as would be the case in the absence of mismatch. In fact, in this extreme example, $\mathsf{mle}_{P,Q}(\gamma)$ and $\mathsf{cmle}_{P,Q}(\gamma)$ not only are not vanishing with increasing $\gamma$ but are increasing without bound because the mismatched distribution has positive mass at 0 and therefore the conditional expectation under it is very close to zero when the channel output has zero occurrences, while the underlying signal value is $1/2$. In this case the incurred loss is very large as $\ell(1/2, \hat{x})$ is unbounded in a neighborhood of 0. So large, in fact, as to cause the overall expected loss to grow without bound with $\gamma$, despite the diminishing probability of observing zero occurrences at the channel output.

# 7 Proofs via Change-of-Measure Formulae

As noted in Section 4, Theorem 4.1 is immediate from Theorem 4.2 once (22) is established. Towards this end note first that

$$D(P_{Y_\gamma} \| Q_{Y_\gamma}) \leq D(P_{X,Y_\gamma} \| Q_{X,Y_\gamma}) \tag{66}$$

$$= D(P \| Q) + D(P_{Y_\gamma | X} \| Q_{Y_\gamma | X} | P) \tag{67}$$

$$= D(P \| Q), \tag{68}$$

where the last equality is due to the fact that, $P$-a.s., $P_{Y_\gamma | X} = Q_{Y_\gamma | X}$ provided $P \ll Q$ (and otherwise $D(P \| Q) = \infty$). The obvious monotonicity of the right hand side of (23) in $\gamma$, put together with (68), implies that the limit in (22) exists and satisfies

$$\lim_{\gamma \to \infty} D(P_{Y_\gamma} \| Q_{Y_\gamma}) \leq D(P \| Q). \tag{69}$$

For the reverse inequality one need merely note that $\gamma^{-1} \cdot Y_\gamma$ converges weakly to $X$ as $\gamma \to \infty$ under both $P$ and $Q$, and thus

$$\liminf_{\gamma \to \infty} D(P_{Y_\gamma} \| Q_{Y_\gamma}) = \liminf_{\gamma \to \infty} D(P_{\gamma^{-1} \cdot Y_\gamma} \| Q_{\gamma^{-1} \cdot Y_\gamma}) \tag{70}$$

$$\geq D(P \| Q), \tag{71}$$

where the inequality follows from the lower semi-continuity of relative entropy under weak convergence [6, Appendix D.3].

We now note that Theorem 4.2 is a direct consequence of Theorem 4.4. Indeed, consider the setting of Theorem 4.4 when $\gamma = 1$ and $X_t \equiv X$ for all $t$ (and all $\omega$) under both measures and note that in this case $E_P[\ell(X_t, E_Q[X_t|Y_1^t])]$ coincides with $\mathsf{mle}_{P,Q}(t)$ from the setting of Theorem 4.2. Thus an application of Theorem 4.4 in this very special case yields Theorem 4.2 (with $T$ playing the role of $\gamma$ in the latter). It remains to prove Theorem 4.3 and Theorem 4.4, to which we dedicate the respective two subsections that follow.

## 7.1 Proof of Theorem 4.3 via Multivariate Point Processes

The main idea is to think of the SNR level as 'time'. In fact, we will use special notation in this subsection, emphasizing this point of view, where $t$ will denote SNR and $z$ will denote the argument for the signal (that elsewhere in this paper is thought of as time). The index set for $z$, i.e. the signal's domain, plays but a secondary role in the analysis, and rather than assuming that it consists of the interval $[0, T]$, we consider it to be a general multidimensional Euclidean set. Our approach is to relate the observation processes $Y_t$ at different SNR $t$ to one another by constructing them as transformations of a *single* observation process that lives in a larger space, namely a multivariate point process driven by the signal. The main tool is the change-of-measure formula for multivariate point processes [16, 17], that is a counterpart of the Girsanov formula for pure jump processes. This formula gives rise to a recursion *in the variable $t$* for the RN derivative, in the form of an integral equation, a key element of the proof. It is here where thinking of SNR as 'time' is useful. In fact, the role played by $t$ (SNR) is analogous to that played by time in the proof of the result on causal estimation provided in the next subsection.

Let $d$ be a positive integer and consider a bounded Borel subset $\mathbb{E}$ of $\mathbb{R}^d$. $\mathbb{E}$ will serve as the domain for the signal process. The given measurable space $(\Omega, \mathcal{F})$ may not be rich enough to support a multivariate point process, so we switch to a new space, and define on it new signal and observation processes, that (in the setting of Subsection 4.2) share with the given signal and observation their joint distribution. Let a measurable space $(\bar{\Omega}, \bar{\mathcal{F}})$ be given, and a random field $X$ over $\mathbb{E}$, taking values in $\mathbb{R}_+$, be defined on it. $X$ is assumed to be measurable as a mapping from $(\bar{\Omega} \times \mathbb{E}, \bar{\mathcal{F}} \otimes \mathcal{E})$ to $(\mathbb{R}_+, \mathcal{R}_+)$, where we write $\mathcal{E}$ for the Borel $\sigma$-field $\mathcal{B}(\mathbb{E})$ and $\mathcal{R}_+$ for $\mathcal{B}(\mathbb{R}_+)$.[1] Also defined on $(\bar{\Omega}, \bar{\mathcal{F}})$ are random variables $Z_n$ and $T_n$, $n \geq 1$, with values in $\mathbb{E}$ and $(0, \infty)$, respectively. $T_n$ are strictly increasing and finite. The sequence $(T_n, Z_n)$ forms a *multivariate point process*, and is characterized by the random measure on $(0, \infty) \times \mathbb{E}$

$$\mu(\omega; dt, dz) = \sum_{n \geq 1} \delta_{(T_n(\omega), Z_n(\omega))}(dt, dz),$$

where $\delta_x$ denotes the unit point mass at $x$. Define a measure on $(\mathbb{E}, \mathcal{E})$ by $\hat{\nu}(B) = \int_B X(z) dz$, $B \in \mathcal{E}$. Let

$$\mathcal{G}_t = \sigma\{\mu((0, s] \times B) : s \leq t, B \in \mathcal{E}\}$$

$$\mathcal{F}_t = \mathcal{F}_0 \vee \mathcal{G}_t$$

where $\mathcal{F}_0 = \sigma\{\hat{\nu}(B) : B \in \mathcal{E}\}$.

Let $\mathbf{P}_0$ (resp., $\mathbf{Q}_0$) be a probability measure on $(\bar{\Omega}, \bar{\mathcal{F}})$ under which $X$ and $\mu$ are mutually independent, $X$ has some given law $P$ (respectively, $Q$), and $\mu$ is a Poisson measure (cf. [17, p. 70]) relative to $\{\mathcal{F}_t\}$ with (deterministic) intensity $\nu_0$ given by

$$\nu_0(dt, dz) = dt dz,$$

where $dt$ and $dz$ denote Lebesgue measures on $(0, \infty)$ and $\mathbb{E}$, respectively. This, by definition, means that $E[\mu(B)] = \nu_0(B)$ for $B \in \mathcal{R}_+ \otimes \mathcal{E}$, and that for any $t \geq 0$, $\mu(\cdot, B)$ is independent of the $\sigma$-field $\mathcal{F}_t$, provided $B \subset (t, \infty) \times \mathbb{E}$. In fact, $\mu$ is the counting measure for what is often called a Poisson point process on $\mathbb{R}_+ \times \mathbb{E}$, and consequently, each $\mu(B)$ is a Poisson r.v. with parameter $\nu_0(B)$, and $\mu(B_i)$ are independent provided $B_i$ are disjoint (see [17, p. 105]). It is assumed that, under both $\mathbf{P}_0$ and $\mathbf{Q}_0$, $X$ is a.s.-bounded between two positive constants.

---

[1] Note that this is consistent with the assumption made in Subsection 4.2 on $X^T$; the more special structure assumed for $X^T$, namely that it has paths in $\mathbb{D}$, is not required here, and is used only in the result on causal estimation.

We invoke an existence result, Theorem 5.2 of [16]. Let $\nu$ denote the random measure on $(\mathbb{E} \times (0, \infty), \mathcal{E} \otimes \mathcal{B}(0, \infty))$

$$\nu(\omega; dt, dz) = \hat{\nu}(\omega; dz)dt = X(\omega; z)dtdz.$$

Then there exists a probability measure $\mathbf{P}$ on $(\bar{\Omega}, \bar{\mathcal{F}})$, $\mathbf{P} \ll \mathbf{P}_0$, under which $X$ has the same law $P$ as under $\mathbf{P}_0$, while $\nu$ is a version of the $\mathcal{F}_t$-predictable projection of $\mu$ for $\mathbf{P}$. We do not define this term here (see [16]), but only note the consequence that, under $\mathbf{P}$, conditionally on $X$, $\mu(B_i)$ are Poisson r.v. with parameters $\nu(B_i)$, i.e. $\int_{B_i} X(z)dtdz$, mutually independent across $i$ provided $B_i$ are disjoint.

Furthermore, [16, Theorem 5.1] (see also [17, Theorem III.5.43]) gives an explicit version of the RN derivative $E_{\mathbf{P}_0}[\frac{d\mathbf{P}}{d\mathbf{P}_0}|\mathcal{F}_t]$ as

$$\Lambda_t = \Big[ \prod_{n:T_n \leq t} X(Z_n) \Big] \exp \int_{\mathbb{E}_t} (1 - X(z))\nu_0(ds, dz),$$

where $\mathbb{E}_t = (0, t] \times \mathbb{E}$ (note that the processes $\alpha$ and $\hat{Y}$ of [16] vanish). This can be expressed as

$$\Lambda_t = \exp \Big\{ \int_{\mathbb{E}_t} \log X(z)\mu(ds, dz) + \int_{\mathbb{E}_t} (1 - X(z))dsdz \Big\}. \tag{72}$$

Let $y_t$ denote the random measure on $\mathbb{E}$, defined by

$$y_t(\omega, B) = \mu((0, t] \times B), \qquad B \in \mathcal{E}.$$

Let $\mathcal{Y}_t = \sigma\{y_t(B) : B \in \mathcal{E}\}$. We can use (72) to calculate the RN derivative between the laws of observation process $\mu$ (respectively, $y$) under both measures, namely $\hat{\Lambda}_t := E_{\mathbf{P}_0}[\Lambda_t|\mathcal{G}_t]$ ($\tilde{\Lambda}_t := E_{\mathbf{P}_0}[\Lambda_t|\mathcal{Y}_t]$).

**Lemma 7.3** *Denote $\hat{X}_t(z) = E_{\mathbf{P}}[X(z)|\mathcal{G}_t]$ and $\tilde{X}_t(z) = E_{\mathbf{P}}[X(z)|\mathcal{Y}_t]$. Then a version of both $\hat{\Lambda}_t$ and $\tilde{\Lambda}_t$ is given by*

$$\hat{\Lambda}_t = \tilde{\Lambda}_t = \exp \Big\{ \int_{\mathbb{E}_t} \log \hat{X}_{s-}(z)\mu(ds, dz) + \int_{\mathbb{E}_t} (1 - \hat{X}_{s-}(z))dsdz \Big\}, \tag{73}$$

*where one can write $\tilde{X}_{s-}(z)$ in place of $\hat{X}_{s-}(z)$ on the r.h.s.*

*Proof:* First note by (72) that $\Lambda_t = \exp\{\int_{\mathbb{E}} \log X(z) y_t(dz) + t \int_{\mathbb{E}} (1-X(z))dz\}$ depends on $\mu$ only through $y_t$. Hence $E_{\mathbf{P}_0}[\Lambda_t|\mathcal{G}_t] = E_{\mathbf{P}_0}[\Lambda_t|\mathcal{Y}_t]$, namely $\hat{\Lambda}_t = \tilde{\Lambda}_t$.

Next, note that $E_{\mathbf{P}}[X(z)|\mathcal{Y}_t] = \frac{E_{\mathbf{P}_0}[X(z)\Lambda_t|\mathcal{Y}_t]}{E_{\mathbf{P}_0}[\Lambda_t|\mathcal{Y}_t]}$. We have just argued that in the denominator one can replace $\mathcal{Y}_t$ by $\mathcal{G}_t$. For a similar reason, the same is true for the numerator. Hence $E_{\mathbf{P}}[X(z)|\mathcal{Y}_t] = E_{\mathbf{P}}[X(z)|\mathcal{G}_t]$, namely $\tilde{X}_t(z) = \hat{X}_t(z)$. To complete the proof it remains to show that $\hat{\Lambda}_t$ is given by the r.h.s. of (73).

Toward this end, we will argue along the lines of Result VI.R8 of [4]. Note that $\Lambda_t$ uniquely solves the integral equation

$$\Lambda_t = 1 + \int_{\mathbb{E}_t} \Lambda_{s-}(X(z) - 1)(\mu(ds, dz) - dsdz). \tag{74}$$

Indeed, this can be seen as a special case of [16, equation (15)], or directly verified by noting that for $t$ between jumps $d\Lambda_t/dt = \Lambda_t \int_{\mathbb{E}}(1 - X(z))dz$, while at times of jump $\int_{\{t\} \times \mathbb{E}} \mu(ds, dz) = 1$ and $\Lambda_t = \Lambda_{t-}X(Z_n) = \Lambda_{t-} \int_{\{t\} \times \mathbb{E}} X(z)\mu(ds, dz)$. Using the relationship between (72) and (74), with $\hat{X}_{s-}(z)$ in place of $X(z)$, it suffices to prove

$$\hat{\Lambda}_t = 1 + \int_{\mathbb{E}_t} \hat{\Lambda}_{s-}(\hat{X}_{s-}(z) - 1)(\mu(ds, dz) - dsdz). \tag{75}$$

By (74) and the independence of $X$ from $\mathcal{G}_t$ under $\mathbf{P}_0$,

$$\hat{\Lambda}_t = 1 + \int_{\mathbb{E}_t} E_{\mathbf{P}_0}[\Lambda_{s-}(X(z) - 1)|\mathcal{G}_s](\mu(ds, dz) - dsdz). \tag{76}$$

Thus (75) will follow once we show equality between the integrands in (75) and (76). Invoking Lemma VI.L6 of [4], showing this amounts to proving

$$E_{\mathbf{P}_0}\Big[\int_{(0,t]\times B} C_s \Lambda_{s-}(X(z)-1)dsdz\Big] = E_{\mathbf{P}_0}\Big[\int_{(0,t]\times B} C_s \hat{\Lambda}_{s-}(\hat{X}_{s-}(z)-1)dsdz\Big] \tag{77}$$

for any $\mathcal{G}_t$-predictable, bounded $C_t$ and $B \in \mathcal{E}$.

Toward showing (77), note first that

$$E_{\mathbf{P}_0}\Big[\int_{(0,t]\times B} \Lambda_{s-}C_s(X(z)-1)dsdz\Big] = E_{\mathbf{P}}\Big[\int_{(0,t]\times B} C_s(X(z)-1)dsdz\Big]. \tag{78}$$

Indeed, using integration by parts, for each $z$,

$$\int_0^t \Lambda_{s-}C_s(X(z)-1)ds = \Lambda_t \int_0^t C_s(X(z)-1)ds - \int_0^t \int_0^s C_u(X(z)-1)d\Lambda_s.$$

Since $\Lambda$ is a martingale under $\mathbf{P}_0$, so is the last term on the right, and

$$E_{\mathbf{P}_0}\Big[\int_{(0,t]\times B} \Lambda_{s-}C_s(X(z)-1)dsdz\Big] = E_{\mathbf{P}_0}\Big[\Lambda_t \int_{(0,t]\times B} C_s(X(z)-1)dsdz\Big].$$

This shows (78).

Next, since under $\mathbf{P}$, the random measures $X(z)dsdz$ and $\hat{X}_{s-}(z)dsdz$ are the predictable projections of $\mu$ w.r.t. $\mathcal{F}_t$ and $\mathcal{G}_t$, respectively (see [16]), we have $E_{\mathbf{P}}\int_{(0,t]\times B} C_s X(z)dsdz = E_{\mathbf{P}}\int_{(0,t]\times B} C_s \hat{X}_{s-}(z)dsdz$. Thus

$$E_{\mathbf{P}}\int_{(0,t]\times B} C_s(X(z)-1)dsdz = E_{\mathbf{P}}\int_{(0,t]\times B} C_s(\hat{X}_{s-}(z)-1)dsdz \tag{79}$$

$$= E_{\mathbf{P}_0}\int_{(0,t]\times B} \Lambda_{s-}C_s(\hat{X}_{s-}(z)-1)dsdz \tag{80}$$

$$= E_{\mathbf{P}_0}\int_{(0,t]\times B} \hat{\Lambda}_{s-}C_s(\hat{X}_{s-}(z)-1)dsdz, \tag{81}$$

where the second equality follows by an argument similar to that leading to (78). Displays (81) and (78) imply (77), which completes the proof. □

We use a convention, analogous to that used elsewhere in this paper, of writing $E_P$ for expectation w.r.t. the joint law of $X$ (signal) and $y_t$ (observation at SNR level $t$) under $\mathbf{P}$. This is legitimate here as well since this joint law is determined by $P$ and the conditional law of $y_t$ as the counting measure of a Poisson point process on $\mathbb{E}$ with intensity $tX$. This is valid also for conditioning, where we will write $\tilde{X}_t^P(z) := E_{\mathbf{P}}[X(z)|\mathcal{Y}_t]$ as $E_P[X(z)|\mathcal{Y}_t]$, and for the law of $y_t$, written $P_{y_t}$. With this notation, applying Lemma 7.3 to both $P$ and $Q$ and noting that the law of $y_t$ under $\mathbf{P}_0$ does not depend on $P$ (hence equal to that under $\mathbf{Q}_0$), a version of $\frac{dP_{y_t}}{dQ_{y_t}}$ is given by

$$\exp\Big\{\int_{\mathbb{E}_t} [\log \tilde{X}_{s-}^P(z) - \log \tilde{X}_{s-}^Q(z)]\mu(ds,dz) - \int_{\mathbb{E}_t} [\tilde{X}_{s-}^P(z) - \tilde{X}_{s-}^Q(z)]dsdz\Big\}.$$

Hence

$$D(P_{y_t}\|Q_{y_t}) = E_P\Big[\int_{\mathbb{E}_t} \log \frac{\tilde{X}_{s-}^P(z)}{\tilde{X}_{s-}^Q(z)}\mu(ds,dz) - \int_{\mathbb{E}_t} (\tilde{X}_{s-}^P(z) - \tilde{X}_{s-}^Q(z))dsdz\Big]. \tag{82}$$

Note by [16] that an integral of the form $\int_{\mathbb{E}_t} R(s,z)(\mu(ds,dz) - \nu(ds,dz))$ forms a martingale under $\mathbf{P}$, provided $R$ is a predictable integrable process. The predictability of $\log \tilde{X}_{s-}^P(z)/\tilde{X}_{s-}^Q(z)$ follows from left continuity, while integrability follows from the boundedness assumptions we put on $X$. Recalling that $\nu(dt,dz)$ is given by $X(z)dtdz$, we have

$$D(P_{y_t}\|Q_{y_t}) = E_P\Big[\int_{\mathbb{E}_t}\Big(X(z)\log\frac{\tilde{X}^P_{s-}(z)}{\tilde{X}^Q_{s-}(z)} - \tilde{X}^P_{s-}(z) + \tilde{X}^Q_{s-}(z)\Big)dsdz\Big] \qquad (83)$$

$$= E_P\Big[\int_{\mathbb{E}_t}\Big(E_P[X(z)|\mathcal{Y}_s]\log\frac{E_P[X(z)|\mathcal{Y}_s]}{E_Q[X(z)|\mathcal{Y}_s]} - E_P[X(z)|\mathcal{Y}_s] + E_Q[X(z)|\mathcal{Y}_s]\Big)dsdz\Big]. \qquad (84)$$

Changing the order of integration, we have thus established the following.

**Theorem 7.6** *For $t \geq 0$,*

$$D(P_{y_t}\|Q_{y_t}) = \int_0^t\int_{\mathbb{E}} E_P\left[\ell\left(E_P[X(z)|\mathcal{Y}_s], E_Q[X(z)|\mathcal{Y}_s]\right)\right]dzds.$$

*Proof of Theorem 4.3:* Specializing to $\mathbb{E} = [0,T]$ and setting $Y_\gamma(z) = y_t([0,z])$, $z \in [0,T]$, where $\gamma = t$, gives for $(X,Y)$ precisely the law indicated in Subsection 4.2 for the signal-observation pair. The result thus follows from Theorem 7.6 as a special case. □

## 7.2 Proof of Theorem 4.4

The tools here parallel those of the previous subsection, but are simpler in that change-of-measure considerations are used only for usual point processes. Naturally, $t$ here will denote time. The treatment cannot be regarded a special case of the one for multivariate point processes, because the signal varies with time (whereas there was no dependence of the signal on SNR). A related (but distinct) calculation is performed in [20, Section 19.5] for mutual information. Here we use a result from Section VI.5 of [4], playing a role similar to that of Lemma 7.3 from the previous subsection.

Let $P, Q \in \mathcal{P}$ be given. Letting $P_\gamma$ (resp. $Q_\gamma$) stand for the distribution of $\gamma \cdot X^T$ when $X^T \sim P$ (resp. $X^T \sim Q$), by the third property in Lemma 2.1 we have

$$\mathsf{cmle}_{P_\gamma,Q_\gamma}(\gamma) = \gamma \cdot \mathsf{cmle}_{P,Q}(\gamma), \qquad (85)$$

implying it suffices to prove Theorem 4.4 assuming $\gamma = 1$. Thus, assuming $X^T = \{X_t, t \in [0,T]\}$ is a non-negative process and that, conditioned on $X^T$, $Y^T$ is a Poisson process of intensity $X^T$, it suffices to prove

$$D(P_{Y^t}\|Q_{Y^t}) = \int_0^t E_P\left[\ell\left(X_s, E_Q[X_s|Y^s]\right) - \ell\left(X_s, E_P[X_s|Y^s]\right)\right]ds \qquad (86)$$

which, in view of Lemma 2.2, is equivalent to

$$D(P_{Y^t}\|Q_{Y^t}) = \int_0^t E_P\left[\ell\left(E_P[X_s|Y^s], E_Q[X_s|Y^s]\right)\right]ds. \qquad (87)$$

Toward proving the above identity, let $\hat{P}, \hat{Q}$ be members of $\hat{\mathcal{P}}_1$ corresponding to $P$ and $Q$, respectively. Also consider an auxiliary probability measure $\hat{P}_0$ on $(\Omega, \mathcal{F})$ under which $X^T$ and $Y^T$ are mutually independent, $X^T$ is distributed $P$, while $Y^T$ is a standard Poisson process. Denote

$$\Lambda_t = \exp\Big\{\int_{(0,t]}\log X_s dY_s + \int_0^t(1-X_s)ds\Big\}. \qquad (88)$$

Then $\Lambda$ uniquely solves the equation

$$\Lambda_t = 1 + \int_{(0,t]}\Lambda_{s-}(X_s - 1)dY_s + \int_0^t \Lambda_s(1-X_s)ds, \qquad (89)$$

and, under $\hat{P}_0$, it is a nonnegative martingale with expected value 1. As a result, $\frac{d\tilde{P}}{d\hat{P}_0} = \Lambda_T$ defines a probability measure $\tilde{P}$ on $(\Omega, \mathcal{F})$. By the analogue of Girsanov's theorem for point processes [20, Theorem 19.4] (cf. also Theorem 19.10 therein; or use the result from the previous subsection for the case where $\mathbb{E}$ is a singleton), the joint law of $(X^T, Y^T)$ under $\tilde{P}$ agrees with that under $\hat{P}$. Consequently, $\Lambda_T$ is a version of the RN derivative between this joint law under $\hat{P}$ and under $\hat{P}_0$, and denoting by $R_{Y^t}$ the law of $Y^t$ under $\hat{P}_0$, writing $E_0$ for $E_{\hat{P}_0}$, we have

$$\frac{dP_{Y^t}}{dR_{Y^t}} = \hat{\Lambda}_t := E_0[\Lambda_t | Y^t], \qquad t \in [0, T]. \tag{90}$$

Result VI(5.6) of [4] (alternatively, an argument along the lines of Lemma 7.3, based on (88), (89)) states

$$\hat{\Lambda}_t = \exp\left(\int_{(0,t]} \log \tilde{X}^P_{s-} dY_s + \int_0^t (1 - \tilde{X}^P_{s-}) ds\right), \tag{91}$$

where $\tilde{X}^P_t = E_P[X_t | Y^t]$. Noting that $R_{Y^t}$ is nothing but a standard Poisson process (in particular, not depending on $P$),

$$\frac{dQ_{Y^t}}{dR_{Y^t}} = \exp\left(\int_{(0,t]} \log \tilde{X}^Q_{s-} dY_s + \int_0^t (1 - \tilde{X}^Q_{s-}) ds\right), \tag{92}$$

whence

$$D(P_{Y^t} \| Q_{Y^t}) = E_P\Big[\int_{(0,t]} \log \frac{\tilde{X}^P_{s-}}{\tilde{X}^Q_{s-}} dY_s - \int_0^t (\tilde{X}^P_{s-} - \tilde{X}^Q_{s-}) ds\Big]. \tag{93}$$

Under $\hat{P}$, $Y_t - \int_0^t X_s ds$ is a martingale, hence so is $\int_0^t Z_s (dY_s - X_s ds)$ for any predictable integrable process $Z$. Thus

$$D(P_{Y^t} \| Q_{Y^t}) = E_P\Big[\int_0^t \Big(X_s \log \frac{\tilde{X}^P_{s-}}{\tilde{X}^Q_{s-}} - \tilde{X}^P_{s-} + \tilde{X}^Q_{s-}\Big) ds\Big] \tag{94}$$

$$= \int_0^t E_P\Big[E_P[X_s | Y^s] \log \frac{E_P[X_s | Y^s]}{E_Q[X_s | Y^s]} - E_P[X_s | Y^s] + E_Q[X_s | Y^s]\Big] ds \tag{95}$$

$$= \int_0^t E_P\left[\ell\left(E_P[X_s | Y^s], E_Q[X_s | Y^s]\right)\right] ds, \tag{96}$$

where the equality in (95) follows by conditioning on $Y^s$. We have thus established (87) and completed the proof.

### 7.3 The Presence of Feedback

An inspection of our proof of Theorem 4.4 reveals that it carries over verbatim to accommodate the presence of feedback. Specifically, the relationship

$$D(P_{Y^T} \| Q_{Y^T}) = \int_0^T E_P\left[\ell\left(X_t, E_Q[X_t | Y^t]\right) - \ell\left(X_t, E_P[X_t | Y^t]\right)\right] dt \tag{97}$$

continues to hold in the case where, under both $P$ and $Q$, the output process $\{Y_t\}_{t \geq 0}$ is a point process which admits the predictable intensity $X_t$, adapted to a suitable filtration. Indeed, the analogue of Girsanov's theorem for point processes that we have employed in the proof accommodates this level of generality. As for the case of non-causal estimation, it is easy to find examples involving the presence of feedback where Equality (25) no longer holds.

## 8 An Alternative Proof Route

We dedicate this section to an outline of an alternative route for proving the main results of this paper, namely Theorems 4.2, 4.3 and 4.4. In this alternative route, rather than starting with the continuous-time setting and then obtaining Theorem 4.2 as a consequence of Theorem 4.4, our starting point is a proof of Theorem 4.2 that is based on first principles. We then extend it to random $n$-vectors, an extension that follows rather directly from the scalar case. Theorem 4.3 is then proven by 'lifting' from finite dimensional vectors to continuous-time processes, establishing first that, when the underlying noise-free signal is piecewise constant, the relative entropy between the true and mismatched distributions of the channel output process coincides with that between the true and mismatched laws of the counts at the edge points of the constancy intervals. This fact allows to appeal to the vector case result to establish Theorem 4.3 for said piecewise constant processes, the general case then following approximation and limiting arguments. Theorem 4.4, on the other hand, can also be proven using Theorem 4.2 as the main building block and an appeal to the chain rule for relative entropy. We omit the details and refer to [33, Section 4-C] for a totally analogous approach taken therein for the Gaussian case.

The merit in this alternative route is in providing further intuition and insight into why the main results hold. We give an elementary proof of Theorem 4.2, and the subsequent results are seen to stem from Theorem 4.2 in a natural way. Because we have already provided rigorous proofs of the main results in the previous section, our goal here is only to outline the main ideas of this alternative proof route, and thus throughout this section we tacitly change orders of limits, summations, integrations, differentiations, etc.

## 8.1 Outline of an Elementary Proof of Theorem 4.2

Noting that

$$D(P_{Y_\gamma}\|Q_{Y_\gamma}) = \sum_{y=0}^{\infty} P(Y_\gamma = y) \log \frac{P(Y_\gamma = y)}{Q(Y_\gamma = y)} \tag{98}$$

$$= \sum_{y=0}^{\infty} \int \frac{e^{-\gamma x}(\gamma x)^y}{y!} dP(x) \cdot \log \frac{\int \frac{e^{-\gamma x}(\gamma x)^y}{y!} dP(x)}{\int \frac{e^{-\gamma x}(\gamma x)^y}{y!} dQ(x)} \tag{99}$$

$$= \sum_{y=0}^{\infty} \int \frac{e^{-\gamma x}(\gamma x)^y}{y!} dP(x) \cdot \log \frac{\int e^{-\gamma x} x^y dP(x)}{\int e^{-\gamma x} x^y dQ(x)}, \tag{100}$$

we have

$$\frac{d}{d\gamma} D(P_{Y_\gamma}\|Q_{Y_\gamma}) \tag{101}$$

$$= \sum_{y=0}^{\infty} \frac{d}{d\gamma} \left[ \int \frac{e^{-\gamma x}(\gamma x)^y}{y!} dP(x) \cdot \log \frac{\int e^{-\gamma x} x^y dP(x)}{\int e^{-\gamma x} x^y dQ(x)} \right] \tag{102}$$

$$= \sum_{y=0}^{\infty} \int \frac{-xe^{-\gamma x}(\gamma x)^y}{y!} dP(x) \cdot \log \frac{\int e^{-\gamma x} x^y dP(x)}{\int e^{-\gamma x} x^y dQ(x)} + \sum_{y=1}^{\infty} \int \frac{xe^{-\gamma x}(\gamma x)^{y-1}}{(y-1)!} dP(x) \cdot \log \frac{\int e^{-\gamma x} x^y dP(x)}{\int e^{-\gamma x} x^y dQ(x)} \tag{103}$$

$$+ \sum_{y=0}^{\infty} \int \frac{e^{-\gamma x}(\gamma x)^y}{y!} dP(x) \cdot \frac{d}{d\gamma} \log \frac{\int e^{-\gamma x} x^y dP(x)}{\int e^{-\gamma x} x^y dQ(x)}. \tag{104}$$

Considering the second sum in (103),

$$\sum_{y=1}^{\infty} \int \frac{xe^{-\gamma x}(\gamma x)^{y-1}}{(y-1)!} dP(x) \cdot \log \frac{\int e^{-\gamma x} x^y dP(x)}{\int e^{-\gamma x} x^y dQ(x)} = \sum_{y=0}^{\infty} \int \frac{xe^{-\gamma x}(\gamma x)^y}{y!} dP(x) \cdot \log \frac{\int e^{-\gamma x} x^{y+1} dP(x)}{\int e^{-\gamma x} x^{y+1} dQ(x)}, \tag{105}$$

and therefore the overall expression in (103) assumes the form

$$\sum_{y=0}^{\infty} \int \frac{-xe^{-\gamma x}(\gamma x)^y}{y!} dP(x) \cdot \log \frac{\int e^{-\gamma x} x^y dP(x)}{\int e^{-\gamma x} x^y dQ(x)} + \sum_{y=1}^{\infty} \int \frac{xe^{-\gamma x}(\gamma x)^{y-1}}{(y-1)!} dP(x) \cdot \log \frac{\int e^{-\gamma x} x^y dP(x)}{\int e^{-\gamma x} x^y dQ(x)} \quad (106)$$

$$= \sum_{y=0}^{\infty} \int \frac{xe^{-\gamma x}(\gamma x)^y}{y!} dP(x) \cdot \log \frac{\int e^{-\gamma x} x^{y+1} dP(x) \int e^{-\gamma x} x^y dQ(x)}{\int e^{-\gamma x} x^{y+1} dQ(x) \int e^{-\gamma x} x^y dP(x)} \quad (107)$$

$$= \sum_{y=0}^{\infty} \int \frac{e^{-\gamma x}(\gamma x)^y}{y!} dP(x) \cdot \frac{\int xe^{-\gamma x}(\gamma x)^y dP(x)}{\int e^{-\gamma x}(\gamma x)^y dP(x)} \cdot \log \frac{\int e^{-\gamma x} x^{y+1} dP(x)}{\int e^{-\gamma x} x^y dP(x)} \frac{\int e^{-\gamma x} x^y dQ(x)}{\int e^{-\gamma x} x^{y+1} dQ(x)} \quad (108)$$

$$\stackrel{(a)}{=} \sum_{y=0}^{\infty} P(Y_\gamma = y) E_P[X|Y_\gamma = y] \log \frac{E_P[X|Y_\gamma = y]}{E_Q[X|Y_\gamma = y]} \quad (109)$$

$$= E_P \left[ E_P[X|Y_\gamma] \log \frac{E_P[X|Y_\gamma]}{E_Q[X|Y_\gamma]} \right], \quad (110)$$

where $(a)$ follows upon noting

$$\frac{\int e^{-\gamma x} x^{y+1} dP(x)}{\int e^{-\gamma x} x^y dP(x)} = \frac{\int x \frac{e^{-\gamma x}(\gamma x)^y}{y!} dP(x)}{\int \frac{e^{-\gamma x}(\gamma x)^y}{y!} dP(x)} = E_P[X|Y_\gamma = y]. \quad (111)$$

Turning to the expression in (104),

$$\sum_{y=0}^{\infty} \int \frac{e^{-\gamma x}(\gamma x)^y}{y!} dP(x) \cdot \frac{d}{d\gamma} \log \frac{\int e^{-\gamma x} x^y dP(x)}{\int e^{-\gamma x} x^y dQ(x)} \quad (112)$$

$$= \sum_{y=0}^{\infty} \int \frac{e^{-\gamma x}(\gamma x)^y}{y!} dP(x) \cdot \frac{\int e^{-\gamma x} x^y dQ(x)}{\int e^{-\gamma x} x^y dP(x)} \cdot \frac{\int -xe^{-\gamma x} x^y dP(x) \int e^{-\gamma x} x^y dQ(x) - \int e^{-\gamma x} x^y dP(x) \int -xe^{-\gamma x} x^y dQ(x)}{\left( \int e^{-\gamma x} x^y dQ(x) \right)^2}$$

$$= \sum_{y=0}^{\infty} \int \frac{e^{-\gamma x}(\gamma x)^y}{y!} dP(x) \cdot \left[ \frac{\int e^{-\gamma x} x^{y+1} dQ(x)}{\int e^{-\gamma x} x^y dQ(x)} - \frac{\int e^{-\gamma x} x^{y+1} dP(x)}{\int e^{-\gamma x} x^y dP(x)} \right] \quad (113)$$

$$= \sum_{y=0}^{\infty} P(Y_\gamma = y) \cdot [E_Q[X|Y_\gamma = y] - E_P[X|Y_\gamma = y]] \quad (114)$$

$$= E_P [E_Q[X|Y_\gamma] - E_P[X|Y_\gamma]]. \quad (115)$$

Thus

$$\frac{d}{d\gamma} D(P_{Y_\gamma} \| Q_{Y_\gamma}) \stackrel{(a)}{=} E_P \left[ E_P[X|Y_\gamma] \log \frac{E_P[X|Y_\gamma]}{E_Q[X|Y_\gamma]} + E_Q[X|Y_\gamma] - E_P[X|Y_\gamma] \right] \quad (116)$$

$$= E_P [\ell (E_P[X|Y_\gamma], E_Q[X|Y_\gamma])] \quad (117)$$

$$\stackrel{(b)}{=} E_P [\ell (X, E_Q[X|Y_\gamma]) - \ell (X, E_P[X|Y_\gamma])] \quad (118)$$

$$= \mathsf{mle}_{P,Q}(\gamma) - \mathsf{mle}_{P,P}(\gamma), \quad (119)$$

where $(a)$ follows from combining (103), (104), (110) and (115), and $(b)$ from the (conditional version of) Lemma 2.2.

## 8.2 Extension to Random Vectors

Let $X^n = (X_1, \ldots, X_n)$ be a random $n$-tuple with non-negative components and, for $\gamma \geq 0$, let $Y_\gamma^n = (Y_{\gamma,1}, \ldots, Y_{\gamma,n})$ be jointly distributed with $X^n$ such that, given $X^n$, the components of $Y_\gamma^n$ are independent with $Y_{\gamma,i}|X^n \sim$ Poisson$(\gamma X_i)$, $1 \leq i \leq n$. If $X^n \sim P$ we denote the distribution of $Y_\gamma^n$ by $P_{Y_\gamma^n}$, conditional expectations by $E_P[X_i|Y_\gamma^n]$, etc. We extend the notation in (20) to this case by

$$\mathsf{mle}_{P,Q}(\gamma) \stackrel{\triangle}{=} E_P \left[ \sum_{i=1}^{n} \ell (X_i, E_Q[X_i|Y_\gamma^n]) \right]. \quad (120)$$

With this extension of the notation, Theorem 4.2 carries over to random vectors essentially verbatim:

**Theorem 8.7** *For any pair of probability laws $P, Q$ governing $X^n$ with components bounded between two positive constants, and for any $\gamma \geq 0$,*

$$D(P_{Y_\gamma^n} \| Q_{Y_\gamma^n}) = \int_0^\gamma [\mathsf{mle}_{P,Q}(\alpha) - \mathsf{mle}_{P,P}(\alpha)] \, d\alpha. \tag{121}$$

*Outline of proof:*[2] For $\beta = (\beta_i, \ldots, \beta_n) \in [0, \infty)^n$, let $Y_\beta^n$ be jointly distributed with $X^n$ such that, given $X^n$, the components of $Y_\beta^n$ are independent and $Y_{\beta,i}|X^n \sim \text{Poisson}(\beta_i X_i)$, $1 \leq i \leq n$. Then

$$\begin{aligned}
\frac{d}{d\gamma} D(P_{Y_\gamma^n} \| Q_{Y_\gamma^n}) &= \sum_{i=1}^n \frac{\partial}{\partial \beta_i} D(P_{Y_\beta^n} \| Q_{Y_\beta^n}) \Big|_{\beta=(\gamma,\ldots,\gamma)} & (122) \\
&\stackrel{(a)}{=} \sum_{i=1}^n \frac{\partial}{\partial \beta_i} \left[ D(P_{Y_\beta^{n\setminus i}} \| Q_{Y_\beta^{n\setminus i}}) + D\left( P_{Y_{\beta,i}|Y_\beta^{n\setminus i}} \,\Big\|\, Q_{Y_{\beta,i}|Y_\beta^{n\setminus i}} \,\Big|\, P_{Y_\beta^{n\setminus i}} \right) \right] \Big|_{\beta=(\gamma,\ldots,\gamma)} & (123) \\
&\stackrel{(b)}{=} \sum_{i=1}^n \frac{\partial}{\partial \beta_i} D\left( P_{Y_{\beta,i}|Y_\beta^{n\setminus i}} \,\Big\|\, Q_{Y_{\beta,i}|Y_\beta^{n\setminus i}} \,\Big|\, P_{Y_\beta^{n\setminus i}} \right) \Big|_{\beta=(\gamma,\ldots,\gamma)} & (124) \\
&= \sum_{i=1}^n \frac{\partial}{\partial \beta_i} \left[ \int D\left( P_{Y_{\beta,i}|y_\beta^{n\setminus i}} \,\Big\|\, Q_{Y_{\beta,i}|y_\beta^{n\setminus i}} \right) dP_{Y_\beta^{n\setminus i}}(y_\beta^{n\setminus i}) \right] \Big|_{\beta=(\gamma,\ldots,\gamma)} & (125) \\
&= \sum_{i=1}^n \int \frac{\partial}{\partial \beta_i} D\left( P_{Y_{\beta,i}|y_\beta^{n\setminus i}} \,\Big\|\, Q_{Y_{\beta,i}|y_\beta^{n\setminus i}} \right) dP_{Y_\beta^{n\setminus i}}(y_\beta^{n\setminus i}) \Big|_{\beta=(\gamma,\ldots,\gamma)} & (126) \\
&\stackrel{(c)}{=} \sum_{i=1}^n \int E_P \left[ \ell(X_i, E_Q[X_i|Y_\beta^n]) - \ell(X_i, E_P[X_i|Y_\beta^n]) \,\Big|\, Y_\beta^{n\setminus i} = y_\beta^{n\setminus i} \right] dP_{Y_\beta^{n\setminus i}}(y_\beta^{n\setminus i}) \Big|_{\beta=(\gamma,\ldots,\gamma)} & (127) \\
&= \sum_{i=1}^n E_P \left[ \ell(X_i, E_Q[X_i|Y_\beta^n]) - \ell(X_i, E_P[X_i|Y_\beta^n]) \right] \Big|_{\beta=(\gamma,\ldots,\gamma)} & (128) \\
&= \sum_{i=1}^n E_P \left[ \ell(X_i, E_Q[X_i|Y_\gamma^n]) - \ell(X_i, E_P[X_i|Y_\gamma^n]) \right] & (129) \\
&= \mathsf{mle}_{P,Q}(\gamma) - \mathsf{mle}_{P,P}(\gamma), & (130)
\end{aligned}$$

where $Y_\beta^{n\setminus i}$ denotes the $n-1$-tuple $(Y_{\beta,1}, \ldots, Y_{\beta,i-1}, Y_{\beta,i+1}, \ldots, Y_{\beta,n})$, (a) is an application of the chain rule of relative entropy (16), (b) is due to the fact that the distributions $P_{Y_\beta^{n\setminus i}}$ and $Q_{Y_\beta^{n\setminus i}}$ do not depend on $\beta_i$, and (c) to an application of Theorem 4.2 with the roles of $P$ and $Q$ played respectively by $P_{X_i|y_\beta^{n\setminus i}}$ and $Q_{X_i|y_\beta^{n\setminus i}}$. $\square$

### 8.3 Outline of Alternative Proof of Theorem 4.3

We start with:

**Lemma 8.4** *Let $X$ be a non-negative random variable, jointly distributed with the pair $(Y_1, Y_2)$ such that, conditionally on $X$, $Y_1$ and $Y_2$ are mutually independent, with both $Y_1|X \sim Poisson(X)$ and $Y_2|X \sim Poisson(X)$. Letting $P$ and $Q$ correspond to two distributions on $X$:*

$$D(P_{Y_1,Y_2} \| Q_{Y_1,Y_2}) = D(P_{Y_1+Y_2} \| Q_{Y_1+Y_2}). \tag{131}$$

---

[2]Theorem 8.7 follows immediately from Theorem 4.3 by specializing the latter to piecewise constant processes. In the outlined proof given here, however, the idea is of course not to rely on Theorem 4.3 and assume only Theorem 4.2, whose elementary proof was outlined in the previous subsection.

*Proof of Lemma 8.4:* Denoting $Z = Y_1 + Y_2$, for any $x \geq 0$, integer $z \geq 0$, and integer $0 \leq y_1 \leq z$

$$P_{Y_1|Z,X}(y_1|z,x) = \frac{P_{Y_1,Z|X}(y_1,z|x)}{P_{Z|X}(z|x)} \tag{132}$$

$$= \frac{\frac{e^{-x}x^{y_1}}{y_1!} \frac{e^{-x}x^{z-y_1}}{(z-y_1)!}}{\frac{e^{-2x}(2x)^z}{z!}} \tag{133}$$

$$= \frac{z!}{y_1!(z-y_1)!2^z} \tag{134}$$

$$= \frac{1}{2^z}\binom{z}{y_1}, \tag{135}$$

which does not depend on the value of $x$ and therefore

$$P_{Y_1|Z}(y_1|z) = \frac{1}{2^z}\binom{z}{y_1}. \tag{136}$$

In particular, the induced conditional distribution of $Y_1$ given $Z$ does not depend on the distribution of $X$ so

$$P_{Y_1|Z} = Q_{Y_1|Z}. \tag{137}$$

Thus

$$D(P_{Y_1,Y_2}\|Q_{Y_1,Y_2}) \stackrel{(a)}{=} D(P_{Y_1,Z}\|Q_{Y_1,Z}) \tag{138}$$

$$= D(P_Z\|Q_Z) + D(P_{Y_1|Z}\|Q_{Y_1|Z}|P_Z) \tag{139}$$

$$\stackrel{(b)}{=} D(P_Z\|Q_Z), \tag{140}$$

where $(a)$ is due to the fact that there is a one-to-one transformation from $(Y_1, Y_2)$ to $(Y_1, Z)$ and $(b)$ is due to (137).
□

Iterating Lemma 8.4 by halving intervals gives

**Corollary 8.4** *Let $X$ be a non-negative random variable $X$, and let $Y^T|X$ be a homogeneous Poisson process of intensity $X$. Letting $P$ and $Q$ correspond to two distributions on $X$,*

$$D(P_{Y^T}\|Q_{Y^T}) = D(P_{Y_T}\|Q_{Y_T}). \tag{141}$$

Equipped with Corollary 8.4, we can start to prove Theorem 4.3 by establishing (25) under the assumption that $X^T$ is piecewise constant both under $P$ and under $Q$, i.e., assume existence of $n$ and a random vector of nonnegative components $A^n = (A_1, \ldots, A_n)$ such that

$$X_t \equiv A_i \quad \text{for all} \quad \frac{i-1}{n}T \leq t < \frac{i}{n}T, \quad \text{and} \quad 1 \leq i \leq n. \tag{142}$$

We use $P$ and $Q$ to denote either the measures governing the continuous-time piecewise-constant signal $X^T$ satisfying (142), or those governing the $n$-dimensional random vector $A^n$. To make the distinction clear, we add the superscript vec in the latter case writing $\mathsf{mle}^{\mathsf{vec}}_{P,Q}(\gamma)$ for $\mathsf{mle}_{P,Q}(\gamma)$ of Section 8.2 (recall (120)) while staying with $\mathsf{mle}_{P,Q}(\gamma)$ for the continuous-time analogue in (24). In the same vein, we let $Y^{\mathsf{vec},n}_\gamma = (Y^{\mathsf{vec}}_{\gamma,1}, \ldots, Y^{\mathsf{vec}}_{\gamma,n})$ denote a random $n$-tuple jointly distributed with $A^n$ such that, given $A^n$, the components of $Y^{\mathsf{vec},n}_\gamma$ are independent with $Y^{\mathsf{vec}}_{\gamma,i}|A^n \sim \text{Poisson}(\gamma A_i)$, while staying with $Y^T_\gamma = \{Y_{\gamma,t}, 0 \leq t \leq T\}$ to denote the process which, conditioned on $X^T$, is a Poisson process of intensity $\gamma X^T$. With this convention, we have

$$\mathsf{mle}_{P,Q}(\gamma) = \frac{T}{n}\mathsf{mle}^{\mathsf{vec}}_{P,Q}\left(\gamma\frac{T}{n}\right) \tag{143}$$

and thus

$$\int_0^\gamma \left[\mathsf{mle}_{P,Q}(\alpha) - \mathsf{mle}_{P,P}(\alpha)\right] d\alpha \stackrel{(a)}{=} \frac{T}{n}\int_0^\gamma \left[\mathsf{mle}^{\mathsf{vec}}_{P,Q}\left(\alpha\frac{T}{n}\right) - \mathsf{mle}^{\mathsf{vec}}_{P,P}\left(\alpha\frac{T}{n}\right)\right]d\alpha \tag{144}$$

$$= \int_0^{\gamma\frac{T}{n}} \left[\mathsf{mle}^{\mathsf{vec}}_{P,Q}(\alpha) - \mathsf{mle}^{\mathsf{vec}}_{P,P}(\alpha)\right] d\alpha \tag{145}$$

$$\stackrel{(b)}{=} D\left(P_{Y^{\mathsf{vec},n}_{\gamma T/n}} \| Q_{Y^{\mathsf{vec},n}_{\gamma T/n}}\right) \tag{146}$$

$$\stackrel{(c)}{=} D(P_{\{Y_{\gamma,iT/n} - Y_{\gamma,(i-1)T/n}\}_{i=1}^n} \| Q_{\{Y_{\gamma,iT/n} - Y_{\gamma,(i-1)T/n}\}_{i=1}^n}) \tag{147}$$

$$\stackrel{(d)}{=} D\left(P_{\{Y_{\gamma,iT/n}\}_{i=1}^n} \| Q_{\{Y_{\gamma,iT/n}\}_{i=1}^n}\right) \tag{148}$$

$$\stackrel{(e)}{=} D(P_{Y^T_\gamma} \| Q_{Y^T_\gamma}), \tag{149}$$

where $(a)$ follows from (143), $(b)$ follows from Theorem 8.7, $(c)$ is due to the fact that $Y^{\mathsf{vec},n}_{\gamma T/n}$ is equal in distribution to $\{Y_{\gamma,iT/n} - Y_{\gamma,(i-1)T/n}\}_{i=1}^n$ regardless of the distribution of the underlying $A^n$, $(d)$ is due to the fact that there is a one-to-one transformation from $\{Y_{\gamma,iT/n} - Y_{\gamma,(i-1)T/n}\}_{i=1}^n$ to $\{Y_{\gamma,iT/n}\}_{i=1}^n$, and $(e)$ is due to an application of Corollary 8.4 on each of the constancy intervals of $X^T$ (and invoking the chain rule of relative entropy).

This concludes the proof for the case where the input is a piecewise constant process of the form in (142). For general processes, given the two process distributions $P$ and $Q$, one considers the induced measures $P^{(n)}$ and $Q^{(n)}$, on the input process $X^{(n),T}$ obtained from the original one via

$$X^{(n)}_t \equiv \frac{2^n}{T}\int_{i2^{-n}T}^{(i+1)2^{-n}T} X_t dt \qquad \text{for } t \in (i2^{-n}T, (i+1)2^{-n}T], \tag{150}$$

a piecewise constant process of the form in (142) for which we have already established the result. Thus, for any $n$,

$$D\left(P^{(n)}_{Y^T_\gamma} \| Q^{(n)}_{Y^T_\gamma}\right) = \int_0^\gamma \left[\mathsf{mle}_{P^{(n)},Q^{(n)}}(\alpha) - \mathsf{mle}_{P^{(n)},P^{(n)}}(\alpha)\right]d\alpha. \tag{151}$$

Standard continuity arguments similar to those in [33, Section IV.C] would now yield

$$D\left(P^{(n)}_{Y^T_\gamma} \| Q^{(n)}_{Y^T_\gamma}\right) \stackrel{n\to\infty}{\longrightarrow} D\left(P_{Y^T_\gamma} \| Q_{Y^T_\gamma}\right) \tag{152}$$

and

$$\int_0^\gamma \left[\mathsf{mle}_{P^{(n)},Q^{(n)}}(\alpha) - \mathsf{mle}_{P^{(n)},P^{(n)}}(\alpha)\right]d\alpha \stackrel{n\to\infty}{\longrightarrow} \int_0^\gamma \left[\mathsf{mle}_{P,Q}(\alpha) - \mathsf{mle}_{P,P}(\alpha)\right]d\alpha, \tag{153}$$

implying (25) when combined with (151).

## 9 Conclusion

Under the right loss function, we find that the Poisson channel exhibits relationships between mutual information, relative entropy, and mismatched estimation loss – for the causal and the non-causal filter– completely paralleling those recently found for the Gaussian channel. For the non-mismatched setting, our findings shed light on the classical continuous-time mutual information relation (43) (cf., e.g., [20]), as well as the recent ones of [14], endowing them with *optimal* estimation interpretations that complete the analogy to Gaussian channel results such as Duncan's formula [7] and the I-MMSE relationship [13].

To what extent our findings can be applied to scenarios involving Poisson channels – analogously to the way their Gaussian counterparts were used in, e.g., [15] for multiuser channels, [22] for analysis of sparse-graph codes, and [30] to establish monotone convergence to a Gaussian limit under relative entropy – remains to explore. It would also be of interest to see whether our results might lead to insight into or improvements on Poisson approximation results, such as those in [2] and references therein. Finally, it would be valuable to understand whether and how the relationships that we now know to hold for the Gaussian and the Poisson channel carry over to other channels. Steps in related directions have been taken recently, establishing that derivatives of information measures with respect to parameters governing the channel induce functionals involving the conditional distribution of the input given the output, cf. e.g. [26, 23, 12, 24]. It remains to be seen whether the latter admit interpretations of operational significance corresponding to optimum or mismatched estimation, and whether these can, in turn, be used to infer insightful relations such as between causal and non-causal estimation.

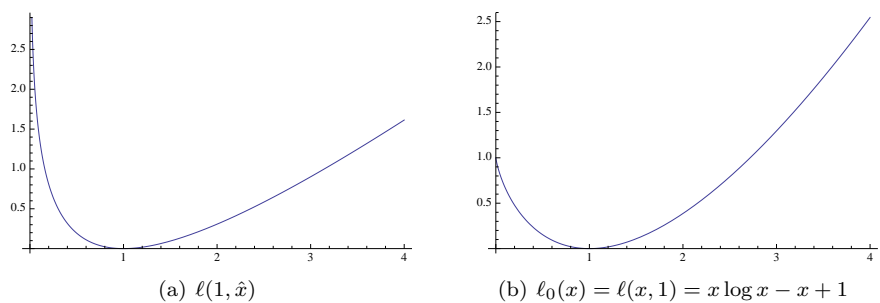

(a) $\ell(1,\hat{x})$  (b) $\ell_0(x) = \ell(x,1) = x\log x - x + 1$

Figure 1: The loss function $\ell$

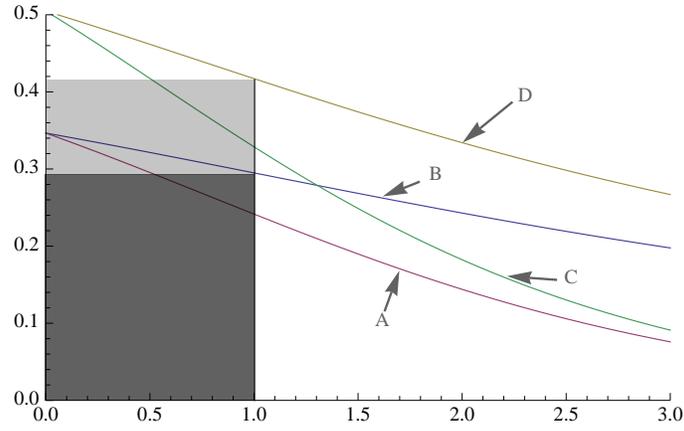

Figure 2: The curves $\mathsf{mle}_{P,P}(\gamma)$, $\mathsf{cmle}_{P,P}(\gamma)$, $\mathsf{mle}_{P,Q}(\gamma)$ and $\mathsf{cmle}_{P,Q}(\gamma)$, marked respectively by A,B,C,D, of the example in Section 6.1, plotted here for $p = 1/2$ and $q = 1/5$.

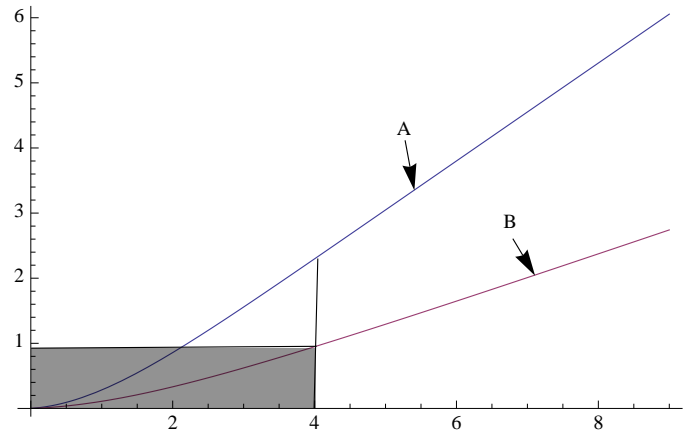

Figure 3: The curves of $\mathsf{mle}_{P,Q}(\gamma)$ and $\mathsf{cmle}_{P,Q}(\gamma)$, as expressed in (62) and (65), marked respectively by A and B. In this example, the mismatched non-causal filter performance is worse than that of the causal one.